\documentclass[aps,pre,onecolumn,showpacs]{revtex4}

\usepackage{epsfig,latexsym}
\usepackage{amsmath,amscd,amssymb}
\newcommand \lan {\langle}
\newcommand \ran {\rangle}
\newcommand \kt {{\tilde \kappa}}

\begin{document}

\title{Thermal denaturation of fluctuating finite DNA chains:\\
the role of bending rigidity in bubble nucleation}
\author{John Palmeri, Manoel Manghi and Nicolas Destainville}

\affiliation{ Laboratoire de Physique Th\'eorique, Universit\'e de
Toulouse, CNRS, 31062 Toulouse, France}


\date{18 September 2007}

\begin{abstract}
Statistical DNA models available in the literature are often
effective models where the base-pair state only (unbroken or broken)
is considered. Because of a decrease by a factor of 30 of the
effective  bending rigidity of a sequence of broken bonds, or
bubble, compared to the double stranded state, the inclusion of the
molecular conformational degrees of freedom in a more general
mesoscopic model is needed. In this paper we do so by presenting a
1D Ising model, which describes the internal base pair states,
coupled to a discrete worm like chain model describing the chain
configurations [J.~Palmeri, M. Manghi, and N. Destainville, Phys.
Rev. Lett. {\bf 99}, {088103} (2007)]. This coupled model is exactly
solved using a transfer matrix technique that presents an analogy
with the path integral treatment of a quantum two-state diatomic
molecule. When the chain fluctuations are integrated out, the
denaturation transition temperature and width emerge naturally as an
explicit function of the model parameters of a well defined
Hamiltonian, revealing that the transition is driven by the
difference in bending (entropy dominated) free energy between bubble
and double-stranded segments. The calculated melting curve (fraction
of open base pairs) is in good agreement with the experimental
melting profile of polydA-polydT and, by inserting the
experimentally known bending rigidities, leads to physically
reasonable values for the bare Ising model parameters. Among the
thermodynamical quantities explicitly calculated within this model
are the internal, structural, and mechanical features of the DNA
molecule, such as bubble correlation length and two distinct chain
persistence lengths. The predicted variation of the
mean-square-radius as a function of temperature leads to a coherent
novel explanation for the experimentally observed thermal viscosity
transition. Finally, the influence of the DNA strand length is
studied in detail, underlining the importance of finite size
effects, even for DNA made of several thousand base pairs. Simple
limiting formul{\ae}, useful for analyzing experiments,  are given
for the fraction of broken base pairs, Ising and chain correlation
functions, effective persistence lengths, and chain
mean-square-radius, all as a function of temperature and DNA length.

\end{abstract}

\maketitle

\section{Introduction}
\label{intro}

The stability of double-stranded DNA (dsDNA) at physiological
temperature is due to the self-assembly of its base pairs:
self-assembly within a same strand {\em via} base-stacking
interactions between neighboring bases; and self-assembly of both
strands {\em via} hydrogen bonds between pairs of complementary
bases.  These interactions, however. are on the order of magnitude
of a few $k_BT$ (thermal energy)~\cite{SantaLucia98,pincet,krueger}
and thermal fluctuations can lead, even at physiological
temperature, to local and transitory unzipping of the double strand
(see {\em e.g.} \cite{wartmont} or the
reviews~\cite{gotoh,lazurkin}). The cooperative opening of a
sequence of consecutive base pairs leads to denaturation bubbles
which are likely to play a role from a biological perspective, since
they may participate in mechanisms such as replication,
transcription or protein binding. For example, it has been
proposed~\cite{Kalosakas04} that transcription start and regulatory
sites could be related to DNA regions which have a higher
probability of promoting bubbles. Indeed, the energy needed to break
an adenine-thymine (A-T) base pair $\sim4 k_BT$, connected by two
hydrogen bonds, is smaller than the energy needed to break a
guanine-cytosine (G-C) one $\sim6 k_BT$ (3 hydrogen
bonds)~\cite{SantaLucia98,pincet}. At the same temperature, A-T rich
sequences present {\em a priori} more bubbles than G-C rich ones,
even though sequence effects on the occurrence of bubbles are more
complex than a simple examination of local A-T base
abundance~\cite{SantaLucia98,Kalosakas04}. In addition to the
sequence, the fraction of denaturation bubbles \textit{in vitro}
naturally depends on temperature, as well as on the ionic strength
of the solution~\cite{gotoh,record}. In particular, the melting
temperature $T_m$, above which bubbles proliferate and the two
strands completely separate, depends on both sequence and ionic
strength. Another parameter that affects the melting temperature is
the length of the double strand~\cite{blake,nelson}. This is not a
purely academic debate because short DNA strands (a few tens of base
pairs) are involved in DNA chip experiments where the hybridization
process is precisely affected by temperature in a way depending on
strand length and sequence (see~\cite{Fiche06} and references
therein).

Although the intracellular unwinding of DNA is due to active and
enzymatic processes by imposing unwinding torsional
stresses~\cite{benham}, the thermally induced denaturation of
purified DNA in solution has led to an intensive study of DNA
thermal
denaturation~\cite{polscher,lazurkin,wartmont,gotoh,wartben}.
Mesoscopic models have been proposed to account for the
thermodynamical properties of denaturation bubbles in DNA. The first
models  were Ising-like two-state models, where the base pairs can
be open or closed (see~\cite{wartmont,gotoh} and references
therein). In the simple base-pair model, the Ising parameters are
the base-pair chemical potential and the so-called cooperativity
parameter, which accounts for the energetic cost of a domain wall.
This type of one-dimensional Ising model  is exactly soluble for
homopolymers and for random sequences~\cite{wartmont}. More
sophisticated effective Ising models, including the 10 kinds of
base-pair doublets, lead to better agreement with experimental
data~\cite{gotoh}. Poland and Scheraga~\cite{polscher}, following
the work of Zimm~\cite{zimm}, included ``loop entropy" in these
models, i.e. the entropic cost of closing large loops formed by
destacked single-stranded DNA (ssDNA) in
bubbles~\cite{polscher,wartmont}. This Poland-Scheraga model is at
the core of the DNA melting simulator MELTSIM algorithm first
developed by Blake \textit{et al.}~\cite{meltsim,blake2}. The
intra-loop and inter-loop self-avoidance corrects the loop
statistics and refines the sharpness of the denaturation
transition~\cite{peliti,carlon}. To get a denaturation transition in
these models, an effective temperature dependent Ising chemical
potential must be inserted by hand. Recently, Peyrard \textit{et
al.} developed non-linear phonon models where the shape of the
interaction potential between base pairs is more precisely taken
into account (see~\cite{peyrard,dauxois,hwa} and the
review~\cite{peyrardreview}). By inserting estimated microscopic
parameters, they have shown that their model can both lead to a
denaturation transition, analogous to interface unbinding, and be
useful for studying bubble dynamics. The predicted transition
temperature is, however, very sensitive to the model parameter
values, and if physically reasonable values are used~\cite{dauxois,
gao, jeon}, $T_m$ appears to be much too high and the transition
width much too large.

More generally, the theoretical study of DNA denaturation can in
principle be tackled on a least four levels of investigation
determined by the amount of detail included, going from (1) quantum
\textit{ab initio} approaches and (2) classical all atom molecular
dynamical simulations~\cite{amber}, through (3) effective mesoscopic
approaches coupling chain conformational and base-pair degrees of
freedom~\cite{orland}, to (4) effective statistical models for
base-pairs alone~(\cite{wartmont,polscher,peyrardreview} and
references therein). In theory, it is possible to move up one step
in this hierarchy by integrating out the subset of the degrees of
freedom that do not appear at the higher level, giving rise to en
effective free energy at each level of description. Our purpose here
is to show, \textit{via} a minimal model for DNA homopolymers, that
if one starts at the third level and integrates out the chain
degrees of freedom, one arrives at a physically coherent level (4)
explanation for the DNA melting transition: a bending free energy
driven denaturation transition emerges naturally due to the entropic
lowering of the energetic barrier for bubble nucleation.  Working at
level (3) also has the added advantage of allowing access to the
statistics of the chain degrees of freedom (effective persistence
length, mean-square-radius, etc.), something that obviously is not
possible at level (4).

To this end, we have recently proposed a model~\cite{prl}, which
considers not only the internal coordinates in terms of Ising spin
variables  describing the open or closed states, but also external
coordinates,  the chain tangent vectors, which determine the chain
configuration and depend sensitively on chain stiffness. Indeed,
ssDNA is two orders of magnitude more flexible than dsDNA at normal
salt concentration. We have shown that this difference in bending
rigidity provides a novel explanation for the bubble mechanism
formation. Further evidence for the importance of this bending
heterogeneity include phenomena such as cyclization, loop formation,
and packaging of DNA into nucleosomes~\cite{Marko04} (where
denaturation bubbles facilitate bending of the otherwise rigid
polymer DNA in structures where it coils up with curvature radii
down to 10~nm, despite a persistence length of dsDNA equal to
50~nm). Our model incorporates precisely this dependence of the
polymer bending rigidity on the state of neighboring base pairs. It
is the discretized version of a continuous model~\cite{prl}, and
couples explicitly an Ising model, describing the internal degrees
of freedom (open or closed), and a Heisenberg or discrete worm like
chain (DWLC) one, accounting for the rotational degrees of freedom
between successive monomers of the DNA chain. Its originality lies
in the fact that the internal 2-state and external bending degrees
of freedom are treated on an equal footing and therefore the
renormalized Ising parameters obtained by integrating out the chain
can be exactly calculated within the scope of the model. The melting
temperature $T_m$ naturally emerges and, together with the
transition width $\Delta T_m$, are explicitly written as a function
of the bending rigidities and strand length, which are
experimentally known, and bare Ising parameters. In addition, the
effective Ising properties (fraction of broken bases and correlation
length) as well as the chain ones (persistence length and mean
square radius) can be computed, allowing in principle direct
comparison with experiments. An important feature of this effective
Ising model is that the end monomers see an effective chemical
potential that is lower than the interior one; this end-interior
asymmetry leads in a natural way to a chain length dependence for
$T_m$, the transition width, and other statistical quantities. A
similar model had been previously proposed by one of us in
2D~\cite{john}, but its application to the study of denaturation
bubbles in dsDNA was not made explicit. Recently, a similar approach
has been considered in the context of the dsDNA stretching
transition~\cite{Nelson03}. The addition in the energy functional of
the term corresponding to the external force prevents an exact
solution of the model in Ref.~\cite{Nelson03} and an approximate
variational scheme had to be implemented.

Beyond dsDNA or dsRNA, our coupled model can be used to describe the
properties of any two-state biopolymer, as soon as the local bending
rigidity depends on the local states. As already mentioned,  the
transition from B- to S-form of dsDNA in force experiments has been
investigated in this framework~\cite{Nelson03}. The helix-coil
transition in poly-peptides can also be described by such a theory
because the $\alpha$-helix configuration is much more rigid than the
random one~\cite{nelson,polscher}.

The present paper is a  detailed account of the results summarized
in a Letter~\cite{prl}. In section~\ref{DSM}, we present the coupled
classical Ising-Heisenberg model, which, as we show here, can be
used to describe DNA thermal denaturation and write the partition
function in terms of a transfer matrix, usual in one-dimensional
statistical systems. We note in passing that the coupled
Ising-Heisenberg model presented here displays a rich array of
behavior and therefore may be of interest in other contexts, such 1D
classical spin chains or 0D quantum rotators describing a diatomic
molecule with internal states.

Because the full transfer matrix method for the coupled model leads
to relatively complex calculations, we first show that the model can
be reduced to two effective Ising models, a path that provides a
great deal of physical insight: indeed, these effective models allow
the calculation of the free energy, as well as the Ising and chain
end-to-end tangent-tangent correlation functions in terms of an
effective Hamiltonian with temperature-dependent Ising parameters. A
detailed solution of the two effective models is provided in
section~\ref{solIsing}. Section~\ref{corr} is devoted to the
calculation of the Ising and chain correlation quantities, using the
effective Ising models, as well as a novel correlator mixing both
Ising and chain variables. The next sections, \ref{TM} and
~\ref{Rg}, present in  detail the full transfer matrix approach,
which leads to the complete calculation of chain correlations and
end-to-end distance; a visual interpretation of the expression for
the 2-point correlation functions leads naturally to  an analogy
with a quantum diatomic molecule. Our theory is compared to
experimental denaturation profiles of synthetic DNA in
section~\ref{DNA} and finite-size effects, which are experimentally
relevant, are thoroughly examined in section~\ref{finite:size}.
Finally, our concluding remarks are given in section~\ref{cl}, where
we also summarize our principal theoretical results of greatest
interest for interpreting experiments. The principal symbols used in
this work are defined and catalogued in Table~\ref{table1} at the
end of the paper.

\section{Discrete Chain Model}

\label{DSM}

We model dsDNA as a discrete chain of $N$ monomers (links), each
monomer can be in one of two different states, U and B, which
denote, respectively, unbroken and broken bonds. The local chain
rigidity depends on the nearby link types. A denaturation bubble is
thus formed by a consecutive sequence of B type monomers. The
chain's conformational properties are determined by the set of $N$
unit link tangent vectors $\{ {\bf{t}}_i :i = 1, \ldots ,N\} $ with
$\|{\bf{t}}_i \| = 1$. For simplicity, the monomer length, $a$, is
taken to be the same for both U and B (for modeling DNA stretching
transitions it is necessary to introduce different monomer
lengths~\cite{nelson, Nelson03}). The position of the end of the
$i^{\rm th}$ link in the 3D embedding space is ${\bf{X}}_i =
{\bf{X}}_0 + a\sum_{j = 1}^i\,{\bf{t}}_j$, where ${\bf{X}}_0 $ is an
arbitrary starting point. The end-to-end vector is ${\bf{R}} =
a\sum_{j = 1}^N\,{\bf{t}}_j$. The  link  states are denoted by the
value of an Ising variable $\sigma _i = \pm 1$ (U or B) associated
with each link. These Ising variables allow us to model a system of
thermally activated defects such as the broken bonds that
proliferate on certain macromolecules like DNA when the temperature
is raised. In this case the temperature-dependent concentration of
broken bonds is controlled by an appropriately defined chemical
potential. Because we are interested here in the new phenomena that
arise due to the coupling between the internal (Ising) and external
(chain conformational) degrees of freedom, we will not attempt to
take into account at the same time the self-avoidance of the chain.

After presenting the model we explain how to calculate the partition
and correlation functions connected with both the internal and
external degrees of freedom for the coupled model. Once we have
these quantities for the coupled Ising-chain system, we will be able
to compare the results for the coupled system with those for the
uncoupled one. We will see that the coupling can strongly modify the
results, namely the average properties of the system at a given
temperature in terms of the average concentration of closed and open
bonds, average mean square chain radius and 2-point correlation
functions.

\begin{figure}[th]
\includegraphics[width=9cm]{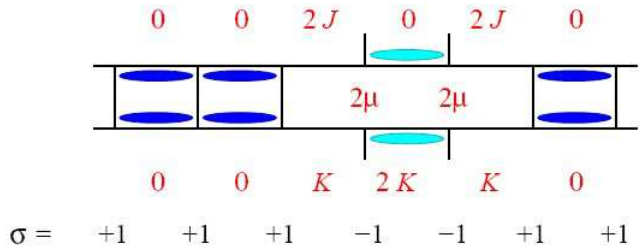}
\caption{Illustration of the different Ising parameters appearing in
the Hamiltonian $\beta H$. Open (closed) base pairs are coded by a
spin $\sigma=+1$ ($-1$). The energies indicate the cost of opening
base-pairs with respect to the ground state where all $\sigma$ are
set to $+1$. The first line shows the cost, $2J$, of a domain wall.
The second line indicates the energy, $2\mu$, required to open a
base pair. The third line gives the difference in stacking energy
between a segment of dsDNA and a denaturated one: dark (light) blue
cigars indicate stacked state in dsDNA (in ssDNA) and the absence of
dots indicates the destacking of adjacent base pairs, which is
already taken into account by the $J$ contribution.} \label{fig1}
\end{figure}

Up to an absolute location in space a state of the chain is given by
the $2N$ variables $\{ \sigma _i ,{\bf{t}}_i \} $. For a chain in 3D
each link vector can be expressed in spherical coordinates as
${\bf{t}}_i  = \left( {\sin(\theta _i ) \cos(\phi _i ),\sin(\theta
_i )\sin(\phi _i ),\cos(\theta _i )} \right)$ and can therefore be
defined by the azimuthal and polar angles $\phi _i $ and $\theta _i
$, denoted together by the solid angle $\Omega _i  = (\theta _i
,\phi _i )$. The energy $H[\sigma _i ,{\bf{t}}_i ]$
 of a state is taken to be
\begin{equation}
H[\sigma _i ,{\bf{t}}_i ] = \sum_{i = 1}^{N - 1}  \, {\tilde \kappa
}_{i + 1,i} (1 - {\bf{t}}_{i + 1}  \cdot {\bf{t}}_i ) - \sum_{i =
1}^{N - 1} \, \left[\tilde J \sigma _{i + 1} \sigma _i +\frac{\tilde
K}2 (\sigma_{i+1} + \sigma _i )\right] - \tilde \mu \sum_{i = 1}^N
\,\sigma_i. \label{H}
\end{equation}
The angle $\gamma _{i,j}$ between two tangent vectors is given by
\begin{equation}
\cos \gamma _{i,j}  = {\bf t}_i  \cdot {\bf t}_j  =
\sin(\theta_i)\sin(\theta _j)\cos(\phi _i - \phi _j ) + \cos(\theta
_i)\cos(\theta _j ). \label{anglecomp}
\end{equation}
The first term in $H$ is the bending energy of a DWLC with a local
rigidity $ {\tilde\kappa }_{i + 1,i}$, having the dimension of
energy, that depends on the neighboring values of the Ising
variables. We have taken, without any loss of generality, the
minimum of the bending energy to be zero independent of the values
of the Ising variables. The second and third terms make up the
energy of the Ising model~\cite{wartmont,gotoh}, $H_{{\rm I}}\equiv
H_{\rm I}(\tilde J,\tilde K,\tilde \mu)$, illustrated in
Fig.~\ref{fig1}.

The term in $\tilde J$ accounts for the local destacking energy
($2\tilde J$) of a domain wall (where $\sigma _i $ passes from one
value to another). The term in $\tilde K$ accounts for the
difference in stacking energy between a segment of dsDNA and of a
denaturation bubble. The last term gives the energy (2$\tilde \mu $)
required to create a link in the state $\sigma _i = -1$ (a B link or
broken bond).  We write a dimensionless Hamiltonian $\beta
H[\sigma_i,\Omega _i ]$, where $\beta = 1/(k_B T)$, which thus
contains the dimensionless parameters $\kappa _{i + 1,i} \equiv
\beta {\tilde \kappa}_{i + 1,i} $, $J \equiv \beta \tilde J$, $K
\equiv \beta \tilde K$, and $\mu\equiv \beta \tilde \mu $. The local
rigidity is
\begin{equation}
\kappa _{i + 1,i}  = \left\{
\begin{array}{ll}
   \kappa _U  & {\rm for \; U-U \; n.n.}  \\
   \kappa _B  & {\rm for\; B-B \; n.n.}  \\
   \kappa _{UB}  & {\rm for\; U-B\; or \; B-U \; n.n.}  \\
\end{array} \right.
\end{equation}
where ``U-U n.n.", etc. denotes nearest neighbor link types. In
terms of the Ising field $\sigma _i $
\begin{equation}
\kappa_{i+1,i}=\frac14(\kappa_U+\kappa_B-2\kappa_{UB})\sigma_{i+1}\sigma_i+
\frac14(\kappa_U-\kappa_B)(\sigma_{i+1}+\sigma_i)+
\frac14(\kappa_U+\kappa_B+2\kappa_{UB}). \label{coupling}
\end{equation}
We identify the B state with two identical non-interacting single
DNA strands, ssDNA, each with a local rigidity equal to
$\kappa_B/2$. From Eqs.~(\ref{H}) and~(\ref{coupling}), the result
of the coupling between Ising and tangent variables can already be
predicted: both the destacking and stacking parameters $\tilde J$
and $\tilde K$ will be modified by the two first terms in
Eq.~(\ref{coupling}) which have exactly the same functional form in
$\sigma_i$, whereas $\tilde\mu$ will remain unchanged. Moreover,
when all the bending rigidities are equal, $\kappa_U
=\kappa_B=\kappa_{UB}$, the two first terms in Eq.~(\ref{coupling})
disappear  and the Hamiltonian~(\ref{H}) decouples into a pure Ising
Hamiltonian and a pure DWLC one (isomorphic to a 1D classical
Heisenberg model for magnetism~\cite{fisher}): $H[\sigma _i
,{\bf{t}}_i ] = H_{\rm DWLC}[{\bf t}_i]+H_{\rm I}[\sigma _i]$. In
the language of magnetism the model studied here is a classical
coupled Heisenberg-Ising spin chain. Although pure effective Ising
models have been used extensively to model helix-coil and
denaturation (melting) transitions in macromolecules, it was
necessary to introduce phenomenologically an effective
temperature-dependent chemical potential to obtain a melting
transition. A key feature of the coupled model used here is that a
melting transition will emerge naturally in the effective Ising
model obtained by integrating out the chain conformational degrees
of freedom.

The quantities that we will use to study the behavior of the coupled
system are the mean of the internal state variable (``magnetization"
in spin language)
\begin{equation}
c \equiv \frac1{N} \sum_{i = 1}^N  \,\sigma _i,
\end{equation}
the local state average, $\langle \sigma _i \rangle$, correlation
functions for the Ising variables,  $\langle{\sigma _{i + r} \sigma
_i } \rangle$, and the chain tangent vectors, $\langle {{\bf{t}}_{i
+ r} \cdot {\bf{t}}_i } \rangle $, and the mean square radius $R
\equiv \langle {\bf R}^2\rangle^{1/2}$. These correlation functions
measure the extent of cooperativity exhibited by the coupled system:
e.g., the size of the B (U) domains below (above) the melting
temperature, and the length scale on which the chain remains rigid.
The concentration of U and B links is given by
\begin{equation}
\varphi _B (N,T) = \frac{1 - \langle c \rangle(N,T)}2= 1 -\varphi _U
(N,T). \label{phidef}
\end{equation}
Once the type of homopolymeric DNA is chosen, the bare Ising
parameters and the chain bending rigidities can be considered fixed,
and therefore $\langle c \rangle$, $\varphi_U$, and $\varphi_B$
become functions of the experimental control parameters, namely
temperature, $T$, and  chain length, $N$. For a pure U chain $
\varphi_U = 1$ and $\varphi _B = 0$; for a pure B chain $ \varphi _U
= 0$ and $ \varphi _B  = 1$. A chain with a finite concentration of
bubbles (B links) will have $ \varphi _B
> 0$ and the melting temperature $T_m < \infty$, if it exists, will
be defined by $ \varphi _U (T_m ) = \varphi _B (T_m ) = 1/2$ . The
equilibrium statistical average of a quantity $\mathcal{O} =
\mathcal{O}[\sigma _i ,\Omega _i ]$ that depends on the fluctuating
degrees of freedom, $[\sigma _i ,\Omega _i ]$, is given by
\begin{equation}
\langle \mathcal{O} \rangle  \equiv \frac{(4\pi )^N}{\mathcal{Z}}
\sum_{\{ \sigma _i  =  \pm 1\} } \,\prod_{i =1}^N  \, \int
\frac{d\Omega _i }{4\pi } \mathcal{O}[\sigma_i,\Omega _i ]e^{ -
\beta H[\sigma _i ,\Omega _i ]} ,
\end{equation}
where
\begin{equation}
\mathcal{Z} = (4\pi )^N \sum_{\{ \sigma _i \} }\, \prod_{i = 1}^N
\,\int \frac{d\Omega _i }{4\pi } e^{ - \beta H[\sigma _i ,\Omega _i
]}
\end{equation}
is the partition function. The partition and correlation functions
for the coupled system can be calculated using transfer matrix
techniques. For example, the partition function can be written as
\begin{equation}
\mathcal{Z} = (4\pi )^N \sum_{\{ \sigma _i \} }  \, \prod_{i = 1}^N
\,\int \frac{d\Omega _i }{4\pi}\langle {V|\sigma _1 } \rangle
\langle \sigma _1 |\hat P(\Omega _1 ,\Omega _2 )|\sigma _2\rangle
\cdots \langle \sigma _{N - 1} |\hat P(\Omega _{N - 1} ,\Omega _N
)|\sigma _N \rangle \langle {\sigma _N |V}\rangle, \label{partfunc}
\end{equation}
where the transfer kernel that appears $N-1$ times in
Eq.~(\ref{partfunc}), is given by
\begin{equation}
\hat P(\Omega _i ,\Omega _{i + 1} ) = \left(
\begin{array}{*{20}c}
   {e^{\kappa _U [\cos(\gamma _{i + 1,i} ) - 1] + J +K+ \mu } } & {e^{\kappa _{UB}
[\cos(\gamma _{i + 1,i} ) - 1] - J} }  \\
   {e^{\kappa _{UB} [\cos(\gamma _{i + 1,i} ) - 1] - J} } & {e^{\kappa _B [\cos(\gamma _{i +
1,i} ) - 1] + J -K- \mu } }  \\
\end{array} \right).
\end{equation}
It is written in the canonical base $|U\rangle=|+1\rangle$ and
$|B\rangle=|-1\rangle$ of the U and B states. The end vector
\begin{equation}
|V\rangle=e^{\mu/2}|U\rangle + e^{-\mu/2}|B\rangle  \label{endv}
\end{equation}
enters in order to take care of the free chain boundary conditions.
Different boundary conditions could be easily handled in a similar
way, for instance for closed (open) ends, $|V\rangle=|U\rangle$
($|B\rangle$) and all the following results not explicitly using
Eq.~(\ref{endv}) remain valid.

Before presenting the full transfer matrix method, we first show
that the partition function and averages of any quantities depending
only on the Ising variables can be obtained by examining the
effective Ising model obtained by integrating over the chain
conformational degrees of freedom (the link tangent vectors) in
Eq.~(\ref{partfunc}). The problem reduces to that of an effective
Ising model with an ``effective free energy" $H_{\rm I, eff}^{(0)}$
containing renormalized parameters. This method works because, for
the coupled Ising-chain system, the rotational symmetry is not
broken (absence of a force term $\propto {\bf t}_i  \cdot {\bf \hat
z}$ in the Hamiltonian~\cite{nelson, Nelson03}). Hence the matrix
obtained by integrating the kernel $\hat P(\Omega _i,\Omega _{i+1})$
in Eq.~(\ref{partfunc}) is the same for any site $i$. We thus are
able to carry out the solid angle integrations in sequential fashion
by using the $(i+1)^{\rm th}$ tangent vector as the polar axis for
the $i^{\rm th}$ solid angle integration. The solid angle integrated
transfer matrix is
\begin{equation}
\hat P_{\rm I, eff}^{(0)}  = \int \frac{{d\Omega _i }}{{4\pi }}\hat
P(\Omega _i ,\Omega _{i + 1} ) = \left( {\begin{array}{*{20}c}
   {e^{ - G_0 (\kappa _U ) + J + K+ \mu } } & {e^{ - G_0 (\kappa _{UB} ) - J} }  \\
   {e^{ - G_0 (\kappa _{UB} ) - J} } & {e^{ - G_0 (\kappa _B ) + J - K - \mu } }  \\
\end{array}} \right)
\label{Pising0}
\end{equation}
where $G_0 (\kappa )$ is the (dimensionless) Helmholtz free energy
of a single joint (two-link) subsystem with rigidity $\kappa$
(either U-U, B-B, U-B or B-U):
\begin{equation}
G_0(\kappa)= -\ln\left\{\int \frac{{d\Omega }}{{4\pi }}\exp[\kappa
(\cos(\theta ) - 1)]\right\} = \kappa  - \ln\left[\frac{\sinh(\kappa
)}{\kappa }\right], \label{G0}
\end{equation}
an increasing function of $\kappa $, varying linearly with $\kappa $
for $\kappa \ll 1$ (high $T$) and as $\ln(2\kappa )$ for $\kappa \gg
1$ (low $T$ entropy dominated regime where the spin wave
approximation for the chain degrees of freedom is valid). The
effective transfer matrix $ \hat P_{\rm I,eff}^{(0)}$ can be written
in Ising form using the renormalized Ising parameters, $J_0$, $K_0$
and the prefactor $\exp(-\Gamma _0)$, all depending on chain
rigidities:
\begin{eqnarray}
\hat P_{\rm I, eff}^{(0)} &=& e^{-\Gamma _0} \left(
{\begin{array}{*{20}c}
   {e^{\mu +K_0  + J_0 } } & {e^{ - J_0 } }  \\
   {e^{ - J_0 } } & {e^{ - \mu -K_0  + J_0 } }  \\
\end{array}} \right)\\
J_0 & \equiv& J - \frac14 \left[G_0 (\kappa _U ) + G_0 (\kappa _B )
-
2G_0 (\kappa _{UB} )\right]\label{J0}\\
K _0  &\equiv& K  - \frac{1}{2}\left[G_0 (\kappa _U ) - G_0 (\kappa
_B)\right]
\label{K0}\\
\Gamma _0  &\equiv& \frac{1}{4}\left[ {G_0 (\kappa _U ) + G_0
(\kappa _B ) + 2G_0 (\kappa _{UB} )} \right].
\end{eqnarray}
In the limit of high temperature, the chain tangents will be
completely  decorrelated and the uninteresting  renormalizations  of
$J$ and $K$ arise solely from $\kappa_{i+1,i}$ (Eq.~\ref{coupling}),
in agreement with Eqs.~(\ref{J0}) and (\ref{K0}), as revealed by the
linear dependence of $G_0 (\kappa)$ on $\kappa$ in this limit. It is
rather in the opposite limit of low temperatures and therefore
strongly correlated chain tangents that the bending entropy driven
denaturation transition arises.

The full partition function, $\mathcal{Z} =\mathcal{Z}_{\rm
I,eff}^{(0)}$, in Ising transfer matrix notation,
\begin{equation}
\mathcal{Z}_{\rm I, eff}^{(0)}= (4\pi )^N \sum_{\{ \sigma _i\} }
\,\langle {V|\sigma _1 } \rangle \langle \sigma _1 |\hat P_{\rm I,
eff}^{(0)} |\sigma _2 \rangle  \cdots \langle \sigma _{N - 1} |\hat
P_{\rm I, eff}^{(0)} |\sigma _N \rangle \langle {\sigma _N |V}
\rangle, \label{Z0}
\end{equation}
can be rewritten explicitly in terms of  an effective Ising free
energy, $H_{\rm I, eff}^{(0)}$:
\begin{equation}
\mathcal{Z}_{\rm I, eff}^{(0)}= (4\pi )^N e^{ - (N - 1)\Gamma _0}
 \sum_{\{ \sigma _i \} }  \,e^{ -\beta H_{\rm I, eff}^{(0)}
[\sigma _i ]}
\end{equation}
where $H_{\rm I, eff}^{(0)}=H_{\rm I}(\tilde J_0,\tilde K_0,\tilde
\mu)$:
\begin{eqnarray}
\beta H_{\rm I, eff}^{(0)} &=&  - J_0 \sum_{i = 1}^{N - 1}\,
\sigma _{i + 1} \sigma _i  - \frac{K_0}2 \sum_{i =1}^{N - 1}\,
(\sigma_{i + 1}+\sigma_i ) - \mu \sum_{i = 1}^N \,\sigma_i \nonumber \\
&=&  - J_0 \sum_{i = 1}^{N - 1}\,\sigma _{i + 1} \sigma _i  -
\frac{L_0}2 \sum_{i =1}^{N - 1}\,(\sigma_{i + 1}+\sigma_i ) -
\frac{\mu}2 (\sigma_1+\sigma_N) \label{Aeff}
\end{eqnarray}
with
\begin{equation}
L_0\equiv \mu +K_0 = \mu +K - \Delta G_0^{UB}/2, \label{L0}
\end{equation}
and $\Delta G_0^{UB} \equiv G_0 (\kappa _U ) - G_0 (\kappa _B)$.
Because $H_{\rm I, eff}^{(0)}$ depends on the temperature, it cannot
be considered as an Ising state energy, but rather as an effective
free energy obtained by integrating out the chain subsystem (cf.
discussion concerning levels of theoretical study in the
Introduction).

When two links open the renormalized stacking energy of the links is
$K_0 $, which is smaller than  $K$ by the difference in bending free
energy between U-B and B-B joints.  The second form of $\beta H_{\rm
I, eff}^{(0)}$ in Eq.~(\ref{Aeff}) shows that the effective chemical
potential of one interior base pair is renormalized to $L_0$. If the
gain in the one link bending free energy in going from U to B,
$\Delta G_0^{UB} $, is greater than the intrinsic energy, $ 2(\tilde
\mu+\tilde K)$, needed to break a closed interior bond, then the
effective interior joint chemical potential $L_0$ can become
negative, signaling a change in stability of U and B states. The end
links $(i = 1,N)$, however, feel a different chemical potential,
$\mu+K_0/2$, which is larger than $L_0$ in the case of interest
($\kappa_U
>\kappa_B $). This end-interior asymmetry, along with the extra
bubble initiation energy due to the second domain wall, are
reflected in the difference between the ``effective free energy"
needed to create an $n$-bubble at a chain end,
\begin{equation}
\beta \Delta G_{{\rm{end}}}^{(n)} = 2J_0  -  K_0  + 2 n L_0
\label{gend}
\end{equation}
and in the chain interior
\begin{equation}
\beta \Delta G_{{\rm{int}}}^{(n)} =  4 J_0  + 2 n L_0, \label{gint}
\end{equation}
which is higher than for an end link by $2 J_0+ K_0$. As will be
confirmed in section~\ref{finite:size}, this difference in effective
free energy suggests that at sufficiently low $T$ bond melting will
begin at the chain ends. Written out in greater detail,
Eq.~(\ref{gint}) leads to
\begin{equation}
\Delta G_{{\rm{int}}}^{(n)} = 4\tilde J_0+2 n k_BT L_0 = 4\tilde J +
2 n (\tilde \mu+\tilde K) - k_BT[G_0(\kappa_U) + G_0(\kappa_B) -2G_0
(\kappa _{UB} )] - n k_BT \Delta G_0^{UB}. \label{gintdet}
\end{equation}
In the uncoupled limit ($\kappa_U = \kappa_B = \kappa_{UB}$) only
the first two (temperature independent) terms survive. Although this
end-interior asymmetry plays no role in the limit of an infinite
chain $(N \to \infty )$, it will play an essential role in
determining how bond melting varies with temperature and bond
location (melting maps) and how the melting temperature varies with
chain size. These finite size effects are  discussed in detail in
section~\ref{finite:size}.

Because the renormalized Ising parameters depend on the chain
parameters, the coupled model does not in general have the same
behavior as the uncoupled one. Indeed, since $G_0 (\kappa )$ is an
increasing function of $\kappa$ and $\kappa_U \gg\kappa_B$ for
dsDNA, if the difference between $\kappa_U$ and $\kappa_B$ is
sufficiently large, then at a certain temperature, $T_m^{\infty } $,
the effective interior bubble chemical potential, $L_0$, can vanish.
Provided that this temperature be sufficiently low for thermal
disorder to be weak and end effects due to the finite size of the
chain to be unimportant, the state of the system will flip from
nearly pure U for $T < T_m^{\infty } $, where $L_0 > 0$, to nearly
pure B for $T > T_m^{\infty} $, where $L _0 < 0$ .  Precisely at $T
= T_m^{\infty} $, there will be, on average, as many closed as open
bonds and $\varphi _{B, \infty} = \varphi _{U, \infty} = 1/2$, where
$\varphi _{B, \infty} (T) \equiv \lim_{N \to \infty} \varphi _{B}
(N, T)$, etc. This transition (strictly speaking a crossover), which
can be extremely sharp under some circumstances (see below), can be
interpreted as a melting or denaturation transition. From the above
analysis, we see clearly that it is the difference in free energy
between U-U and B-B joints, $\Delta G_0^{UB}$, that drives the
melting transition. We will see below, moreover, that if $\kappa _U
$ and $\kappa _B $ are much greater than one, then the spin wave
approximation is valid, and the entropy term in $\Delta G_0^{UB} $
dominates. We will see in section~\ref{DNA} that this is actually
the case for real DNA.

By calculating the average of the product of the cosines of the $N-
1$ angles, $\prod_i \cos(\gamma _{i + 1,i})$ and using the same
technique used above for $\mathcal{Z}_{\rm I, eff}^{(0)}$, we can
define another effective model, but now with partition function
\begin{equation}
\mathcal{Z}_{\rm I, eff}^{(1)}=\mathcal{Z}\left\langle\prod_{i
=1}^{N-1}\,{\bf t}_{i + 1}\cdot{\bf t}_i \right\rangle.
\end{equation}
This expression  can be written in Ising transfer matrix form as
\begin{equation}
\mathcal{Z}_{\rm I, eff}^{(1)}= (4\pi )^N \sum_{\{\sigma_i\} }
\,\langle {V|\sigma _1 } \rangle \langle \sigma_1 |\hat P_{\rm I,
eff}^{(1)} |\sigma_2 \rangle  \cdots \langle {\sigma _{N - 1} |\hat
P_{\rm I, eff}^{(1)} |\sigma _N } \rangle \langle {\sigma _N |V}
\rangle, \label{Z1}
\end{equation}
where
\begin{eqnarray}
\hat P_{\rm I, eff}^{(1)}  &=& \int \frac{d\Omega _i }{4\pi} \cos
(\gamma _{i + 1,i}) \hat P(\Omega _i ,\Omega _{i + 1} ) =e^{- \Gamma
_1} \left( {\begin{array}{*{20}c}
   {e^{\mu +K_1  + J_1 } } & {e^{ - J_1 } }  \\
   {e^{ - J_1 } } & {e^{ - \mu -K_1  + J_1 } }  \\
\end{array}} \right) \label{Pising1} \\
J_1  &\equiv& J - \frac{1}{4}[G_1 (\kappa _U ) + G_1 (\kappa _B ) -
2G_1 (\kappa _{UB} )]\\
K_1  &\equiv& \mu  - \frac{1}{2}[G_1 (\kappa _U ) - G_1 (\kappa _B
)]\\
\Gamma _1  &\equiv& \frac{1}{4}.
\end{eqnarray}
The function $G_1(\kappa) = -\ln\{\int \frac{d\Omega}{4\pi}
\cos(\theta)\exp[\kappa (\cos(\theta ) - 1)]\}$ is related to the
tangent-tangent correlation function between two adjacent monomers
(isolated 2-link sub-system) with rigidity $\kappa$:
\begin{equation}
\left\langle {{\bf{t}}_1  \cdot {\bf{t}}_2 } \right\rangle _{{\rm{2
- link}}}  = \left\langle {\cos(\theta )} \right\rangle _{{\rm{2 -
link}}}  = \exp\left[- G_1 (\kappa ) + G_0 (\kappa ) \right]
=u(\kappa )=\coth(\kappa ) - 1/\kappa, \label{G1}
\end{equation}
which is the Langevin function~\cite{joyce}. It increases with
$\kappa$, varying as $\kappa /3$ for $\kappa \ll 1$ and as $1 -
1/\kappa $ for $\kappa \gg 1$. This asymptotic behavior corresponds
to $G_1 (\kappa ) - G_0(\kappa )$ varying as $\ln(3/\kappa )$ for
$\kappa \ll 1$ and as 1/$\kappa $ for large $\kappa $. For a pure
chain (U or B) $\langle\cos(\theta)\rangle _{2 - {\rm link}}$ is
equal to the nearest-neighbor tangent correlation function
$\langle{\bf t}_{i + 1} \cdot {\bf t}_i \rangle =\exp(-1/\xi_p)$
with persistence length
\begin{equation}
\xi _p (\kappa) =-1/\ln[u(k)]= [G_1(\kappa )-G_0 (\kappa )]^{-1}.
\label{xi}
\end{equation}
In the high $\kappa$ (spin wave) approximation, we obtain $\xi _p
\simeq \kappa$, as expected.

The partition function can be rewritten explicitly in terms of an
effective Ising free energy, $H_{\rm I, eff}^{(1)}$:
\begin{equation}
\mathcal{Z}_{\rm I, eff}^{(1)}  = (4\pi)^N e^{-(N - 1)\Gamma _1}\;
\sum_{\{ \sigma _i\}}\,e^{-\beta H_{\rm I, eff}^{(1)} [\sigma _i ]},
\end{equation}
where
\begin{equation}
\beta H_{\rm I, eff}^{(1)}  =  - \sum_{i = 1}^{N - 1}
\,\left[J_1\sigma _{i + 1} \sigma _i  + \frac{L_1}2 (\sigma _{i + 1}
+ \sigma _i )\right] - \frac{\mu}2 (\sigma _1  + \sigma _N ) .
\end{equation}

By repeatedly using the vector identity $({\bf a}\cdot {\bf c})({\bf
b} \cdot {\bf d}) = ({\bf a} \cdot {\bf d})({\bf b} \cdot {\bf
c})-({\bf a} \times {\bf b})\cdot ({\bf c} \times {\bf d})$ for
${\bf{a}} \cdot {\bf{b}} = {\bf{t}}_{i + 1}\cdot {\bf{t}}_i$, etc.,
along with the property that averages of cross products, $\langle
{\bf{t}}_{i + 1}\times {\bf{t}}_i\rangle$, are zero (and that
$\left|{{\bf{t}}_i } \right|^2 = 1$), the partition function
$\mathcal{Z}_{\rm I, eff}^{(1)}$ can be written as the product of
the end-end tangent-tangent correlation function, $\langle
{\bf{t}}_1 \cdot {\bf{t}}_N \rangle$, and the effective Ising
partition function, $\mathcal{Z}_{\rm I, eff}^{(0)}$:
\begin{equation}
\mathcal{Z}_{\rm I, eff}^{(1)} = \left\langle {\bf{t}}_1  \cdot
{\bf{t}}_N \right\rangle \mathcal{Z}_{\rm I, eff}^{(0)}.
\end{equation}
When $\kappa _U  = \kappa _B  = \kappa _{UB}  = \kappa$, we recover
the pure chain tangent-tangent correlation function:
\begin{equation}
\langle {\bf{t}}_1  \cdot {\bf{t}}_N \rangle = \frac{
\mathcal{Z}_{\rm I, eff}^{(1)}}{\mathcal{Z}_{\rm I, eff}^{(0)}} \to
\left\{\frac{\exp[-G_1(\kappa )]}{\exp[-G_0 (\kappa )]}\right\}^{N -
1}  = \exp \left[-(N-1)/\xi_p(\kappa)\right].
\end{equation}

Coming back to the difference in (dimensionless) free energy $\Delta
G_0^{UB}$, we can show that at room temperature it is dominated by
its entropic part. Indeed, $G_0(\kappa)$ can be split into an
average (dimensionless) energy and average entropy contribution,
$G_0(\kappa) =  E_0(\kappa)-S_0(\kappa) /k_B $, where
$E_0(\kappa)=\beta\,\partial G_0/\partial \beta  = \kappa\,\partial
G_0 (\kappa)/\partial \kappa$. Hence, the average energy of a
two-link system can be written in terms of $G_0 $ and $G_1$:
\begin{equation}
E_0 (\kappa ) = \kappa \left[1-\exp\left(-G_1(\kappa) +
G_0(\kappa)\right)\right] = \kappa \left[1-\exp\left(-1/\xi_p
(\kappa)\right)\right] = \kappa \left[1-u(\kappa)\right],
\end{equation}
and therefore $\Delta E_0^{UB}\equiv E_0(\kappa
_U)-E_0(\kappa_B)=(\kappa _U -\kappa _B) -[\kappa_U u(\kappa _U
)-\kappa_B u(\kappa _B)]$. Because the function $u(\kappa)$ tends to
1, $\Delta E_0^{UB}\to 0$, and therefore for temperatures low enough
for the spin wave approximation to be valid for both U-U and B-B
links, we see that $\Delta G_0^{UB} \simeq-\Delta S_0^{UB}/k_B$.
Indeed in this approximation, the Hamiltonian is Gaussian and
equipartition of energy occurs: $\langle\tilde
E\rangle\sim\beta^{-1} \Rightarrow\langle
E\rangle=\beta\langle\tilde E\rangle\sim1$. In this case the melting
transition, if it exists, will be driven overwhelmingly by the
difference in entropy between U-U and B-B joints. As done for
$\Delta G_0^{UB} $, the difference in (dimensionless) free energy
$\Delta G_1^{UB}$ can be split into an average (dimensionless)
energy and average entropy contribution using the same formul{\ae}
as above. If $L_1= 0$ at a certain temperature, $T_1^{\infty }$,
then we can expect another type of ``melting" transition, now driven
by the free energy difference $\Delta G_1^{UB}\equiv
G_1(\kappa_U)-G_1(\kappa_B)$. We return to this point below.

It should be noticed that the above processus could in principle be
carried on (with increasing difficulty) to calculate higher order
correlation functions quantities. Hence these effective Ising models
give information on multi-point tangent-tangent correlation
functions. In the next section we obtain the solutions to the two
effective Ising models.

\section{Solution of the two Ising models}
\label{solIsing}

The effective Ising partition and correlation functions can be
obtained using well-known Ising transfer matrix
techniques~\cite{nelson}. In order to treat in parallel
$\mathcal{Z}_{\rm I, eff}^{(0)}$ and $\mathcal{Z}_{\rm I,
eff}^{(1)}$, we introduce the index $l=0,1$ and compute the
associated partition function $\mathcal{Z}_{\rm I, eff}^{(l)}$. This
index will be useful in Section~\ref{TM} where we introduce the
transfer matrix approach. We need the eigenvectors and eigenvalues
of the transfer matrices: $\hat P_{\rm I, eff}^{(l)} | {\psi ^{(l)}
}\rangle  = \lambda_l | \psi ^{(l)} \rangle$, where the expressions
for $ \hat P_{\rm I, eff}^{(l)}$ are given in
Eqs.~(\ref{Pising0},\ref{Pising1}). The eigenvalues are
\begin{equation}
\lambda _{l, \pm }  = e^{J_l  - \Gamma _l } \left\{ {\cosh(L_l ) \pm
\left[ \sinh^2 (L_l ) + e^{ - 4J_l } \right]^{1/2} } \right\}
\label{vpl}
\end{equation}
and they obey the inequality $ \lambda _{l, + }  > \lambda _{l, -}$.
The two orthonormal eigenvectors, $| \psi^{(l)}\rangle$, are
\begin{equation}
|l,+\rangle  = \frac1{\sqrt{2 \gamma _l}e^{J_l}} \left( a_l
|U\rangle + a_l^{-1} |B\rangle\right) \quad  \mathrm{and} \quad
|l,-\rangle = \frac1{\sqrt{2 \gamma _l}e^{J_l}}  \left(a_l^{-1}
|U\rangle - a_l |B\rangle\right), \label{vectpl}
\end{equation}
where
\begin{equation}
\gamma _l = \left[\sinh^2 (L_l ) + e^{-4J_l} \right]^{1/2}  \quad
\mathrm{and}  \quad  a_l  = e^{J_l } \left[\sinh (L_l) + \gamma _l
\right]^{1/2} \label{al}.
\end{equation}

The transfer matrices can be expanded in terms of the eigenvectors
\begin{equation}
\hat P_{\rm I, eff}^{(l)}  = \sum_{\tau  =  \pm } {\lambda _{l,\tau
} \left| {l,\tau } \right\rangle } \left\langle {l,\tau } \right|.
\end{equation}
Using the decomposition of the Ising $(2 \times 2)$
 identity matrix, $\hat I_I  = \sum_{\tau} | l,\tau\rangle \langle l,\tau |$
 the orthonormality of the eigenvectors $| l,\tau\rangle$ for each value of
 $l$ [i.e., $\langle l,\tau | l,\tau '\rangle  = \delta _{\tau ,\tau '}$],
 and the forms Eqs.~(\ref{Z0},\ref{Z1}) for the effective Ising partition
 functions, we then find
\begin{equation}
\mathcal{Z}_{\rm I, eff}^{(l)}  = (4\pi )^N \left\langle V
\right|\left[\hat P_{\rm I, eff}^{(l)}\right]^{N-1} \left| V
\right\rangle  = (4\pi )^N \sum_{\tau=\pm} \lambda _{l,\tau }^{N-1}
\langle V|{l,\tau}\rangle^2. \label{Zl}
\end{equation}
The full partition is given by $ \mathcal{Z} = \mathcal{Z}_{\rm I,
eff}^{(0)}$. The matrix elements, $\langle V|{l,\tau}\rangle$,
entering Eq.~(\ref{Zl}) can be obtained explicitly from
Eqs.~(\ref{endv}), (\ref{vectpl}), and (\ref{al}).

When $\kappa _U = \kappa _B = \kappa _{UB}$, the system decouples,
but the pure Ising model with temperature-independent parameters,
$\tilde J$, $\tilde K$ and $\tilde \mu$, exhibits neither a second
order phase transition at a finite temperature in zero ``field"
($\tilde \mu +\tilde K= 0$), nor a melting transition at a finite
temperature for $\tilde \mu+\tilde K> 0$. Indeed, there can be no
melting transition because the inequality $\varphi _B  < 1/2$ holds
over the whole temperature range. At low temperatures, ${\sinh(\mu
+K) \gg e^{ - 2J} }$, and the system is ordered with $\varphi _U
\simeq 1$ and $\varphi _B \simeq 0$. As the temperature is raised,
denaturation bubbles are thermally excited, with a cross-over when
$\sinh(\mu +K) \simeq e^{ - 2J}$, or roughly $\beta' =(k_B T')^{-1}
\simeq( {\tilde \mu +\tilde K + 2\tilde J})^{ - 1} $. At higher
temperatures ($\sinh(\mu+K) \ll e^{-2J}$), the average
``magnetization" monotonously approaches a completely thermally
disordered state with $\langle c\rangle= 0$ and $\varphi _U =\varphi
_B= 1/2$. Note that this regime is not reached for DNA since as we
will see below, $T'\simeq 8\,T_m$ and the model is certainly no
longer valid for such high temperatures. By contrast, the coupled
Ising-chain model will exhibit a very different behavior, with a
finite temperature melting transition. In the following we
implicitly assume that all temperatures of interest obey $T \ll T'$.

Although the eigenvectors $| l,\tau \rangle$ are orthogonal in $\tau
$ for the same value of $l$, this is  not necessarily the case for
different values of $l$ (as we will see below, this is a consequence
of the difference in rotational symmetry). In general, depending on
the values of the temperature-dependent effective Ising parameters,
$L_l $ and $J_l $, the eigenvectors are complicated mixtures of the
canonical basis states, $| U \rangle$ and $| B\rangle$. These
eigenvectors and their corresponding eigenvalues can, however, be
simplified in two important limits:
\begin{itemize}
    \item For sufficiently low or high temperatures, below  or above
    the transition temperature, $T_l^{\infty} $ (at which $L_l $ vanishes),
    the inequality $\sinh^2 (L_l) \gg e^{-4J_l}$ is obeyed (the experimental
    melting temperature for infinite chains is thus $T_m^{\infty} \equiv
    T_0^{\infty}$). As a consequence,
    the off-diagonal (domain-wall or ``tunneling") terms in $\hat P_{\rm I, eff}^{(l)}$
    can be neglected and the eigenvectors reduce asymptotically  to the canonical ones,
    with the mapping depending on the sign of $L_l$: $| l, + \rangle  \simeq | U \rangle$
    and $| {l, - }\rangle \simeq -| B\rangle$ for $L_l  > 0$
    and $| {l, - } \rangle \simeq | U \rangle$ and $| {l,+}\rangle \simeq | B\rangle$
    for $L_l< 0$. In this limit of strong cooperativity,
    the eigenvalues reduce to
\begin{equation}
\lambda _{l, \pm }  \simeq \exp \left( {J_l  \pm \left| {L_l}
\right| - \Gamma _l } \right) \quad \mathrm{for} \quad \sinh^2 (L_l)
\gg e^{-4J_l}
\end{equation}
and therefore the pure U state is strongly favored if $L_l> 0$ and
the B state if $L_l < 0$, because
\begin{equation}
\frac{{\lambda _{l, + } }}{{\lambda _{l, - } }} \simeq \exp \left(
{2\left| {L_l } \right|} \right) \quad \mathrm{for} \quad \sinh^2
(L_l) \gg e^{-4J_l}.
\end{equation}
By introducing the following eigenvalues for pure $U$ and pure $B$
\begin{eqnarray}
   \lambda_{l,U} &\equiv& \exp [J +\mu-G_l(\kappa_U)]  \\
   \lambda_{l,B} &\equiv& \exp [J -\mu-G_l(\kappa_B)]
\end{eqnarray}
and using the definitions of  $ J_l ,L_l$, and $ \Gamma _l$, these
limiting forms for the eigenvalues can be further simplified
($\sinh^2 (L_l) \gg e^{-4J_l}$):
\begin{equation}
\begin{array}{l}
   \lambda_{l,+} \simeq \lambda_{l,U} \\
   \lambda_{l,-} \simeq  \lambda_{l,B}  \\
\end{array}
\quad \mathrm{for} \quad L_l  > 0 \quad \mathrm{and} \quad
\begin{array}{l}
   \lambda_{l,+}  \simeq  \lambda_{l,B} \\
   \lambda_{l,-}  \simeq  \lambda_{l,U} \\
\end{array}
\quad \mathrm{for} \quad L_l  > 0. \label{limitingf}
\end{equation}

    \item For temperatures at or very near the transition temperature,
    however, the opposite inequality $\sinh^2 (L_l) \ll e^{-4J_l}$ holds
    and $L_l$ can be set to zero in Eqs.~(\ref{vpl})--(\ref{al}):
    the eigenvectors then reduce to symmetric and anti-symmetric,
    superpositions of the canonical basis vectors, $|U\rangle$ and $|B\rangle$:
\begin{equation}
\left| {l, \pm } \right\rangle  \simeq \frac1{\sqrt2}\left( {\left|
U \right\rangle  \pm \left| B \right\rangle } \right)  \quad
\mathrm{for} \quad \sinh^2 (L_l) \ll e^{-4J_l}
\end{equation}
with eigenvalues
\begin{equation}
\lambda _{l,\pm}  \simeq e^{ - \Gamma _l } (e^{J_l}\pm e^{-J_l})
 \quad \mathrm{for} \quad \sinh^2 (L_l) \ll e^{-4J_l}.
 \label{limitingvp}
\end{equation}
In this limit of weak cooperativity the eigenvalue ratio is
approximately $\lambda _{l,+}/\lambda_{l,-}\simeq \coth(J_l)$ and
the behavior of the system is dominated by the domain walls. The
average behavior for large $N$, which is governed by the ground
symmetrical state, shows vanishing average for the mean of the Ising
state variable (or magnetization in spin language) (for $l=0$ and $N
\to \infty$, $\varphi _{U} \simeq \varphi _{B} \simeq 1/2$, since
there is no spontaneous second order phase transition for the 1D
zero field Ising model).
\end{itemize}

\begin{figure}[t]
\includegraphics[height=6cm]{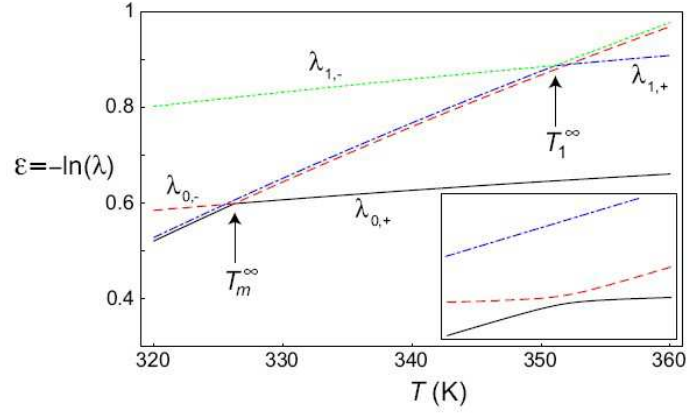}
\caption{``Energies" $\varepsilon_{l, \pm} = -\ln (\lambda_{l,
\pm})$ for $l=0,1$ in the quantum formalism (related to the
Landau-Zener problem) vs. temperature (for parameter values used in
section~\ref{DNA} to fit experimental melting data:   $\tilde
\mu=4.46$~kJ/mol, $\tilde J=9.13$~kJ/mol and $\tilde K=0$). We
observe that far from the two transition temperatures, the
eigenvalues reduce to the limiting forms Eq.~(\ref{limitingf}). The
inset is a zoom close to $T_m^\infty$ showing the level repulsion
between the branches ($0,\pm$). A similar level repulsion occurs
near $T_1^\infty$ between the branches ($1,\pm$).} \label{fig2}
\end{figure}

The exact results for the eigenvalues  and eigenvectors interpolate
smoothly between the above simplified results obtained far from and
close to $T_l^{\infty} $ (Fig.~\ref{fig2}).

Using Eq.~(\ref{endv}), these   limiting forms can  be used to
obtain simple approximations for the following important matrix
elements :
\begin{equation}
 \langle V |0,+ \rangle  =
\left\{
\begin{array}{ll}
   e^{\mu/2},  &  T <  T_m^\infty   \\
   \sqrt2 \cosh(\mu/2),  &  T =  T_m^\infty  \\
   e^{-\mu/2},  &  T >  T_m^\infty \\
\end{array} \right. \label{matelp},
\end{equation}
\begin{equation}
 \langle V |0,- \rangle =
\left\{
\begin{array}{ll}
   -e^{-\mu/2},  &  T <  T_m^\infty   \\
   \sqrt2 \sinh(\mu/2),  &  T =  T_m^\infty  \\
   e^{\mu/2},  &  T >  T_m^\infty. \\
\end{array} \right. \label{matelm}
\end{equation}
The limiting forms for $\langle V |0,- \rangle$ reveal that this
matrix element is negative for low temperature and positive at
$T_m^\infty$ and therefore must pass through zero at a temperature
$T^*$  lower than $T_m^\infty$. This special temperature will be
studied in detail in the section concerning finite size effects
(Section~\ref{finite:size}).

If a melting transition exists at a finite temperature,
$T_m^\infty$, then $L_0$ will go from a positive value below
$T_m^\infty$, through zero at the transition, then to a negative
value above $T_m^\infty$. This temperature dependence for $L_0$ is
similar to the time dependence of the uncoupled energy levels in the
Landau-Zener problem, a quantum 2-state dipole system in an
electrical field varying linearly with time~\cite{landauzener}. It
is not surprising, therefore, that the 2 branches for the
``adiabatic" states of the Landau-Zener problem are equivalent, as
the time $t$ varies from $-\infty$ to $+\infty$, to the
``eigenenergies", $\varepsilon_{l, \pm}  \equiv -\ln (\lambda_{l,
\pm})$, of the states $| {l, \pm}\rangle $ presented above, as the
temperature varies from below the transition to above (with the same
type of limiting behavior near and far from the transition
temperature, $T=T_l^\infty$, equivalent to $t = 0$ in the quantum
Landau-Zener problem, see Fig.~\ref{fig2}). As in the quantum
mechanics of diatomic molecules, eigenenergies possessing the same
rotational symmetry (same $l$) cannot cross (level repulsion),
although they do  reach a point of closest approach at $T_m^\infty$.
As an illustration we show in Fig.~\ref{fig2}  the Landau-Zener
diagrams, $\varepsilon_{l, \pm} (T)$ for $l=0,1$ and observe level
repulsion near $T=T_m^\infty$ and $T=T_1^\infty$ for states with
same $l$ and the possibility of level crossing for states with
different values of $l$.

From the full partition function, we define the (dimensionless)
Helmholtz free energy per Ising variable of the coupled system,
$F=-N^{-1}\ln \mathcal{Z}$. The average value of the Ising state
variable (or ``magnetization") can then be obtained for finite
chains using $\langle c\rangle  =- \partial F/\partial \mu$, from
which $\varphi_U$ and $\varphi _B$ can be deduced. The expression
for $\langle c \rangle$ simplifies in  the limit $N \to \infty $,
because only the largest of the eigenvalues, $\lambda_{0,+}$,
entering in the $l=0$ effective Ising partition function survives:
\begin{equation}
\langle c \rangle\underset{N\to\infty}{\to} \left\langle c
\right\rangle_\infty \equiv - \frac{\partial f}{\partial \mu} =
\frac{\sinh(L_0)}{[\sinh^2(L_0) + e^{-4J_0}]^{1/2}} \label{c}
\end{equation}
where $f={\lim}_{N\to\infty} F = -\ln \lambda _{0,+ }$.
Equation~(\ref{c}) can then be used to find $\varphi_{U, \infty}$
and $\varphi _{B, \infty}$.

If $L_0$ vanishes at a temperature, $T_m^{\infty}$,  low enough for
the $e^{-4J_0}$ term in the denominator to be sufficiently small,
then the system will undergo a sharp melting transition: $\langle c
\rangle_\infty$ will jump sharply from $+1$ for $T < T_m^{\infty}$
(pure U state) to $-1$ for $T > T_m^{\infty}$ (pure B state). The
size of $e^{-4J_0}$ term  in Eq.(\ref{c})  will determine the width
of the transition region,
\begin{equation}
\Delta T_m^{\infty}  \equiv 2 \left| \frac{\partial \left\langle c
\right\rangle_\infty }{\partial T} \right |_{T_m^{\infty}}^{-1}
\simeq \frac{2\,k_B [T_m^{\infty}]^2}{\tilde \mu} \exp
[-2\,J_0(T_m^{\infty})] \label{deltaTm}
\end{equation}
which is exponentially small in $J_0(T_m^{\infty})$ when
$J_0(T_m^{\infty}) \gg 1$.

In a similar manner, from the effective partition function,
$\mathcal{Z}_{\rm I,eff}^{(1)}$, we can define the free energy per
Ising spin, $F^{(1)} = - N^{ - 1} \ln \mathcal{Z}_{\rm I, eff}^{(1)}
$. A quantity analogous to the average Ising ``magnetization",
$\langle c \rangle$, for this partition function can then be
obtained for finite chains using
\begin{equation}
\langle c^{(1)} \rangle  =  - \frac{\partial F^{(1)}} {\partial \mu}
= \frac{{\left\langle {c\,({\bf{t}}_1  \cdot {\bf{t}}_N )}
\right\rangle }}{{\left\langle {\bf{t}}_1  \cdot {\bf{t}}_N
\right\rangle }},
\end{equation}
from which $\varphi _U^{(1)}=(1 + \langle c^{(1)}\rangle)/2$ and
$\varphi _B^{(1)} $ can be obtained. From the definition of $\langle
c^{(1)} \rangle$, we see that it is a mixed correlation function
that describes how the average system internal state, $c$, is
correlated with its external state (chain configuration)
\textit{via} the end-end tangent-tangent correlation function. In
the limit of an infinite chain $\langle c^{(1)}\rangle$ reduces to
an expression analogous to Eq.~(\ref{c}):
\begin{equation}
\langle c^{(1)} \rangle \underset{N\to\infty}{\to} \langle c^{(1)}
\rangle_\infty =-\frac{{\partial f^{(1)} }}{{\partial \mu }} =
\frac{{\sinh(L_1 )}}{{ {\left[ {\sinh^2(L_1)+ e^{ -4J_1 } }
\right]}^{1/2} }},\label{cinf}
\end{equation}
where $ f^{(1)} = {\lim }_{N \to \infty } F^{(1)}$.  This limiting
formula is similar to the one obtained for $\langle c \rangle$ in
Eq.~(\ref{c}), which implies that if $L_1$ vanishes at a
temperature, $T_1^{\infty} $,  low enough for the $e^{ - 4J_1 }$
term in the denominator to be sufficiently small, then the system
can undergo a second type of ``melting" transition: $\langle c^{(1)}
\rangle$ will jump sharply from $+1$ for $T < T_1^\infty$ (pure $
c^{(1)}$ ``U" state) to $-1$ for $T > T_1^\infty$ (pure $c^{(1)}$
``B" state). For $T_m^\infty<T<T_1^\infty$, $\langle c^{(1)}
\rangle\simeq+1$ while $\langle c \rangle\simeq-1$. This implies
that in this temperature range $\langle c\,({\bf t}_1 \cdot {\bf
t}_N ) \rangle \simeq  - \langle c \rangle \langle {\bf t}_1 \cdot
{\bf t}_N \rangle$. This counter-intuitive result is another
manifestation of the non-trivial coupling between internal and
external degrees of freedom. The value of the $e^{ - 4J_1 }$ term
determines again the width of the transition region:
\begin{equation}
\Delta T_1^{\infty}  \equiv 2 \left| \frac{\partial \langle
c^{(1)}\rangle_\infty}{\partial T} \right|_{T_1^{\infty}}^{ - 1}
\simeq \frac{2\,k_B[T_1^{\infty}]^2}{\tilde \mu} \exp[- 2\,
J_1(T_1^{\infty })].
\end{equation}

\section{Ising state variable--Ising and chain correlation functions}
\label{corr}

The average value of the local Ising variable $\left\langle { \sigma
_i } \right\rangle$ and the  2-point $\left\langle {\sigma _{i + r}
\sigma _i } \right\rangle $ correlation function can be calculated
by starting from the expression Eq.~(\ref{partfunc}) for the
partition function and using the property that an insertion of a
term $\sigma_j$ in the sum of products is equivalent to the
insertion of the Pauli matrix in the canonical basis,
\begin{equation}
\hat \sigma _z  = \left(
\begin{array}{*{20}c}
  1 & 0  \\
  0 & { - 1}  \\
\end{array} \right)  \label{pauliz}
\end{equation}
at the $j^{\rm th}$ position in the product of transfer matrices
defining the partition function, Eq.~(\ref{Zl}). This comes from
$\lan\sigma|\hat \sigma_z|\sigma'\ran=
\sigma\delta_{\sigma,\sigma'}$ and the equality
\begin{equation}
\sigma_j \langle\sigma _j| \hat P_{\rm I,eff}^{(0)} |\sigma_{j+1}
\rangle = \sum_{\sigma=\pm 1}\,\langle\sigma_j |\hat \sigma _z
|\sigma \rangle \langle\sigma | \hat P_{\rm
I,eff}^{(0)}|\sigma_{j+1} \rangle = \langle\sigma_j| \hat \sigma_z
\cdot \hat P_{\rm I,eff}^{(0)}|\sigma_{j+1} \rangle.
\end{equation}
We then find
\begin{eqnarray}
\langle {\sigma _i }\rangle  &=& \frac{(4\pi)^N}{\mathcal{Z}}
\left\langle V \right|\left[ {\hat P_{\rm I,eff}^{(0)} } \right]^{i
- 1} \hat \sigma _z \left[ {\hat P_{\rm I,eff}^{(0)} } \right]^{N -
i} \left|
V \right\rangle\\
\langle {\sigma _{i + r} \sigma _i }\rangle  &=&
\frac{(4\pi)^N}{\mathcal{Z}} \left\langle V \right|\left[ {\hat
P_{\rm I,eff}^{(0)} } \right]^{i - 1} \hat \sigma _z \left[ {\hat
P_{\rm I,eff}^{(0)} } \right]^r \hat \sigma_z \left[ {\hat P_{\rm
I,eff}^{(0)} } \right]^{N - r - i} \left| V \right\rangle.
\end{eqnarray}
By the same method used for reducing the partition function, we
finally obtain
\begin{eqnarray}
\left\langle {\sigma _i } \right\rangle
&=&\frac{(4\pi)^N}{\mathcal{Z}} \sum_{\tau _1,\tau _2 }  \langle V
|0,\tau_2 \rangle \lambda _{0,\tau _2 }^{i - 1} \langle 0,\tau _2
|\hat \sigma _z | 0,\tau _1 \rangle \lambda _{0,\tau _1 }^{N - i}
\langle {0,\tau _1 } | V
\rangle \label{isingave}\\
\left\langle {\sigma _{i + r} \sigma _i } \right\rangle &=&
\frac{(4\pi)^N}{\mathcal{Z}}  \sum_{\tau _1,\tau _2,\tau _3 }
\langle V |0,\tau_3 \rangle \lambda _{0,\tau _3 }^{i - 1} \langle
0,\tau _3 |\hat \sigma _z | 0,\tau _2 \rangle \lambda _{0,\tau _2
}^r \langle 0,\tau _2|\hat \sigma _z | 0,\tau _1\rangle \lambda
_{0,\tau _1 }^{N - r - i} \langle {0,\tau _1 } | V \rangle
\label{isingcorr}
\end{eqnarray}
with $\tau _i  =  \pm $. The matrix elements appearing in the above
expressions can be found explicitly using Eqs.~(\ref{endv}),
(\ref{vectpl}), and (\ref{al}).

The Pauli matrix $ \hat \sigma _z$, which can be interpreted as a
quantum mechanical dipole moment operator, is diagonal in the
canonical basis, $| U \rangle= |+1\rangle $ and $| B \rangle=|
-1\rangle $ [see Eq.~(\ref{pauliz})], but not necessarily in the
basis that diagonalizes $ \hat P_{\rm I,eff}^{(l)}$. Indeed, in the
$l = 0$ basis we have
\begin{equation}
\hat \sigma _z ^{(0)} = \left( \begin{array}{*{20}c}
   \langle c\rangle_\infty & (1- \langle c\rangle_\infty)^{1/2}  \\
  (1- \langle c\rangle_\infty)^{1/2} & { -  \langle c\rangle_\infty}  \\
\end{array} \right).\label{sigmaz0}
\end{equation}
On the one hand, the transfer matrix $ \hat P_{\rm I,eff}^{(l)}$
mixes the canonical basis states, which explains the complicated
representation of the effective Ising partition function,
Eqs.~(\ref{Z0}) and (\ref{Z1}), state variable average,
Eq.~(\ref{isingave}), and  correlation function,
Eq.~(\ref{isingcorr}) in this basis. On the other hand, in the basis
that diagonalizes the transfer matrix, the propagation between
measurements is simple (no mixing), but now, in general, the
``dipole" measurement process, corresponding to $ \hat \sigma _z$,
mixes such states. By directly summing Eq.~(\ref{isingave}) over $i$
and using the matrix elements of Eq.~(\ref{sigmaz0}), we can
calculate $\langle c\rangle (N, T)$:
\begin{equation}
\langle c\rangle  (N, T) = \langle c\rangle_\infty \left[ 1 -
\frac{2 R_V^{2}}{R_V^{2}+ e^{(N-1)/\xi_I} } \right] + \frac{2 R_V
\sqrt{1-\langle c\rangle_\infty^2} \left[ 1- e^{-(N-1)/\xi_I}
\right]}{N \left[ 1+R_V^2 e^{-(N-1)/\xi_I}  \right]
\left[1-e^{-1/\xi_I} \right] }, \label{cn}
\end{equation}
where
\begin{equation}
R_V \equiv \frac{\langle V |0,- \rangle }{ \langle V |0,+ \rangle}
\label{RV}
\end{equation}
and $\xi_{I}$ is the Ising correlation length
\begin{equation}
\xi_{I}=1/\ln(\lambda_{0,+}/\lambda_{0,-}),\label{ising},
\end{equation}
the typical size of minority B (U) domains below (above) $T_m$.
Although $\langle V |0,- \rangle$ can be positive or negative (and
even zero), $\langle V |0,+ \rangle$ is for physical reasons
strictly positive, because both $|V \rangle$ and $|0,+ \rangle$ are
linear combinations of the canonical basis states with strictly
positive  coefficients of proportionality [see Eqs.~(\ref{vectpl}),
(\ref{al}), (\ref{matelp}), and (\ref{matelm})]. The ratio $R_V$
(which can therefore be negative, zero, or positive) is thus always
well defined.

The above expression, Eq.~(\ref{isingcorr}), for $\langle\sigma _{i
+ r} \sigma _i\rangle$ can be interpreted, using ``path integral"
imagery, as a quantum mechanical measurement process over an
imaginary time period of $N$ steps of duration $\delta$. This
interpretation is based on the 1D classical Ising representation of
the partition function of a 0D quantum 2 state
system~\cite{chandler}: the transfer matrix becomes the quantum
propagator, $\hat P_{{\rm I},{\rm eff}}^{(0)}  \leftrightarrow \exp
(-\delta \hat H/\hbar)$, where $\hat H$ is the quantum Hamiltonian,
and the eigenvalues, $\lambda $, of the transfer matrix are related
to $\varepsilon $ , the energy eigenvalues of the Hamiltonian via
$\varepsilon  \leftrightarrow  -\ln \lambda $. In general the two
states for the quantum system are coupled by a non-nonzero tunneling
amplitude, which corresponds to the off-diagonal (domain-wall) terms
of the transfer matrix. For instance, following
Eq.~(\ref{isingcorr}), the system is prepared in the initial state
$| V \rangle $ and evolves $N - r - i$ time steps under the dynamics
determined by the propagator $\hat P_{\rm I,eff}^{(0)}$, until a
measurement of the dipole moment is performed, determined by the
action of $\hat \sigma _z$. The state that comes out of the
measurement then evolves $r$ time steps until a second  measurement
of the dipole moment is performed. The state that comes out then
evolves $i - 1$ further time steps. The correlation function
$\langle\sigma_{i + r} \sigma_i \rangle$ is thus the normalized
amplitude that the system returns to the initial state $| V\rangle $
at the end of this double measurement process.

In the limit $N \to \infty $, the results
Eqs.~(\ref{isingave})-(\ref{isingcorr}) simplify because we keep
only the leading order terms (largest eigenvalues) which sets
$\tau_1= +$:
\begin{eqnarray}
\left\langle {\sigma _i } \right\rangle &\underset{N\to\infty}{\to}&
\langle c \rangle_\infty +  R_V  ( 1 - \langle c
\rangle_\infty^2)^{1/2}
\exp [ -(i-1)/\xi_I] \label{isingaveinf}\\
\langle\sigma_{i + r} \sigma _i\rangle &\underset{N\to\infty}{\to}&
\sum_{\tau_3,\tau_2}\frac{\langle V |{0,\tau_3} \rangle}{\langle V|
{0,+}\rangle} \left(\frac{\lambda_{0,\tau_3}}{\lambda
_{0,+}}\right)^{i - 1} \langle {0,\tau_3} | \hat \sigma_z |
{0,\tau_2} \rangle \left(\frac{\lambda_{0,\tau_2}}{\lambda_{0, +
}}\right)^r \langle {0,\tau_2} |\hat \sigma _z | {0, + }\rangle,
\label{isingcorrinfN}
\end{eqnarray}
where we have used $\left\langle {\sigma _i } \right\rangle_\infty
\equiv \langle c \rangle_\infty$.  Using the above results we obtain
the limiting form for $\langle c\rangle (N, T)$ when $N \to \infty$:
\begin{equation}
\langle c\rangle (N, T) \underset{N\to\infty}{\to}\langle
c\rangle_\infty + \frac{2}{N} \frac{R_V \sqrt{1-\langle
c\rangle_\infty^2} }{
 1-e^{-1/\xi_I} }, \label{cninf}
\end{equation}

In the double limit $N,i \to \infty$, meaning that we ignore the
influence of end-monomers,  expressions Eqs.~(\ref{isingaveinf}) and
(\ref{isingcorrinfN}) reduce to the simpler cyclic boundary
condition forms:
\begin{eqnarray}
\langle \sigma _i\rangle &\underset{N,i\to\infty}{\to}&  \langle c
\rangle_\infty \label{sigmaisimple} \\
\langle\sigma _{i + r} \sigma _i\rangle
&\underset{N,i\to\infty}{\to}& \langle c\rangle_\infty^2 +
\left(1-\langle c\rangle_\infty^2\right) \exp(-r/\xi_{I}).
\label{isingcorrsimple}
\end{eqnarray}
Using the limiting values obtained earlier, Eq.~(\ref{limitingvp}),
we find that at the melting temperature, $T_m^{\infty}$, $\xi_I =
-1/\ln[\coth(J_0)]$. When  $e^{-2J_0(T_m^{\infty})} \ll 1$, $\xi_I
(T_m^{\infty}) \simeq e^{2J_0(T_m^{\infty})}/2\gg 1$. We shall see
in section~\ref{DNA} that $\xi_{I}$ can be extremely large, but
finite, at $T_m^{\infty}$, where it reaches its maximum value.
Moving away from $T_m^{\infty}$ in both directions, we find that
$\xi_{I} \simeq 1/(2|L_0|)$ decreases as $|T-T_m^{\infty}|$
increases. When $\xi_I (T_m^{\infty}) \gg 1$,  the width of the the
transition, Eq.~(\ref{deltaTm}), can be rewritten as $\Delta
T_m^{\infty} \simeq k_B \left[ T_m^{\infty} \right]^2
/[\tilde{\mu}\xi_{I}(T_m^{\infty})]$. Because the system is
translationally invariant in the limit $N,i \to \infty $ and
$\langle (\sigma _{i + r}  -\langle\sigma_{i +r}\rangle)(\sigma_i
-\langle\sigma_i\rangle) \rangle =\langle\sigma _{i + r} \sigma
_i\rangle - \langle\sigma _i \rangle^2$,
Eqs.~(\ref{sigmaisimple})-(\ref{isingcorrsimple}) show that for
$|\langle c\rangle_\infty| \neq 1$ the Ising correlation length
measures the range of correlations between spatially
separated deviations of the local Ising spin from the average value.\\

Using the same Ising model transfer matrix techniques employed above
for the partition function, we can calculate the chain end-end
tangent-tangent correlation function, $\left\langle {\bf{t}}_1 \cdot
{\bf{t}}_N  \right\rangle$, which is related to the effective
partitions functions $\mathcal{Z}_{\rm I, eff}^{(l)}$, $l = 0,1$:
\begin{equation}
\langle {\bf t}_1  \cdot {\bf t}_N\rangle  = \frac{\mathcal{Z}_{\rm
I, eff}^{(1)}}{\mathcal{Z}_{\rm I, eff}^{(0)}} = \frac{\sum_{\tau}
\lambda_{1,\tau }^{N-1}  \langle V | 1,\tau\rangle^2}{\sum_{\tau}
\lambda_{0,\tau }^{N-1} \langle V | {0,\tau } \rangle^2} = \frac {
\langle V |{1, + }\rangle^2 \exp \left[ { - (N - 1)/\xi _{1, + }^p }
\right] + \langle V |{1, - }\rangle^2 \exp \left[ { - (N - 1)/\xi
_{1, - }^p } \right] } {\langle V |{0, + }\rangle^2  + \langle V
|{0, - }\rangle^2 \exp \left[ { - (N - 1)/\xi _I } \right]}
\label{tangentNcorr}
\end{equation}
where the chain persistence lengths are defined by
\begin{equation}
\xi_{1,\pm}^p =
1/\ln(\lambda_{0,+}/\lambda_{1,\pm}).\label{persists}
\end{equation}
This result indicates clearly that in general $\langle {\bf t}_1
\cdot {\bf t}_N\rangle$ depends on three distinct characteristic
lengths: $\xi_{I}$ and $\xi_{1,\pm}^p$.

In order to better understand the physical content of
Eq.~(\ref{tangentNcorr})  (and later results), it is useful to
derive simplified limiting forms  for the  matrix elements and
persistence lengths appearing therein. Using the same technique
employed to obtain the expressions for $\langle V |0,\pm \rangle$
shown in Eqs.~(\ref{matelp}) and (\ref{matelm}), simple limiting
forms can be derived for $\langle V |1,\pm \rangle$ in the
temperature range of experimental interest, $T < T_1^\infty$,
leading to $\langle V |1,+ \rangle \simeq e^{\mu/2}$ and $\langle V
|1,- \rangle \simeq -e^{-\mu/2}$.

Using the limiting values obtained earlier for the eigenvalues,
Eq.~(\ref{limitingvp}), we find the following limiting forms for the
two chain persistence lengths:
\begin{equation}
 1/\xi _{1, + }^p  \simeq
\left\{
\begin{array}{lr}
   1/\xi _{U}^p,  &  T <  T_m^\infty  \\
   1/\xi _{I} + 1/\xi _{U}^p,  &  T_m^\infty <T<T_1^\infty  \\
\end{array} \right.
\quad \mathrm{and} \quad
 1/\xi _{1, - }^p  \simeq
\left\{
\begin{array}{lr}
   1/\xi _{I} + 1/\xi _{B}^p, &  T <  T_m^\infty   \\
   1/\xi _{B}^p,  &  T_m^\infty <T<T_1^\infty. \\
\end{array} \right.  \label{limpers}
\end{equation}

The limiting forms for $\xi_{1,+}^p$ in the range  $T_m^\infty
<T<T_1^\infty$ and  $\xi_{1,-}^p$ in the range $T<T_m^\infty$ have a
simple physical explanation: the effective persistence lengths, $\xi
_{1, \pm }^p$, of minority domains (B, or $-$, below $T_m$ and U, or
$+$, above $T_m$) tend to the typical minority domain size, $\xi
_{I}$, when these domains behave as rigid rods ($\xi _{U}^p \gg \xi
_{I}$ for minority U domains and $\xi _{B}^p \gg \xi _{I}$ for
minority B domains).

Is is interesting to note that the various correlation lengths can
be identified with differences between the eigenenergies appearing
in the Landau-Zener diagram (Fig.~\ref{fig2}): $\xi _{I} =
1/(\varepsilon_{0, -} - \varepsilon_{0, +})$, $\xi _{1, + }^p =
1/(\varepsilon_{1, +} - \varepsilon_{0, +})$, and $\xi _{1, - }^p =
1/(\varepsilon_{1, -} - \varepsilon_{0, +})$. Hence we already
observe in this diagram that $\xi_I$ reaches its maximum at
$T_m^\infty$ which is the point of closest approach of the branches
($0,\pm$). In the limit of large $N$ the expression for $\langle
{\bf t}_1 \cdot {\bf t}_N\rangle$ substantially simplifies and
depends on only one persistence length, $\xi _{1, + }^p$:
\begin{equation}
\langle {\bf t}_1 \cdot {\bf t}_N\rangle \underset{N\to\infty}{\to}
\frac{\langle V |{1, + }\rangle^2 }{\langle V |{0, +}\rangle^2} \exp
\left[ { - (N - 1)/\xi _{1, + }^p } \right].
\label{tangentNcorrNinf}
\end{equation}

The value of $N$ for which the limiting form
Eq.~(\ref{tangentNcorrNinf}) starts to be  a good approximation to
Eq.~(\ref{tangentNcorr}) depends on the temperature via the weights,
$\langle V |{l, + }\rangle^2$ and the characteristic lengths,
$\xi_{I}$ and $\xi_{1,\pm}^p$.

\section{Full transfer matrix approach}
\label{TM}

To calculate the general chain tangent-tangent correlation function,
$ \langle {\bf{t}}_i \cdot {\bf{t}}_{i + r} \rangle$, for the
coupled model, we need to introduce the more powerful (and more
abstract) transfer kernel method. This method will also shed
additional light on the origin of the effective Ising models
obtained above by first integrating out the chain degrees of
freedom. To calculate the partition and correlation functions using
this method, we need to solve a spinor eigenvalue problem in order
to find the eigenfunctions and eigenvalues of the transfer kernel:
$\hat P| {\hat \Psi } \rangle  = \lambda | {\hat \Psi } \rangle$, or
more explicitly
\begin{equation}
\sum_{\sigma ' =  \pm 1}  \,\int {\frac{{d\Omega '}}{{4\pi }} {\hat
P}_{\sigma ,\sigma '} (\Omega ,\Omega ')\Psi _{\sigma '} (\Omega ')}
\, = \lambda \Psi _\sigma  (\Omega ), \label{eigenvalpb}
\end{equation}
where
\begin{equation}
| {\hat \Psi}\rangle  =  \Psi_{+1}(\Omega )|U\rangle + \Psi_{-1}
(\Omega) |B\rangle .
\end{equation}

For the pure Ising model the eigenvalues and eigenvectors can be
labeled by the index $\tau = \pm $. For the pure chain model with
rigidity $\kappa $ and no applied stretching force (like the 1D
classical Heisenberg model in zero field) the eigenfunctions, $\psi
_{lm} (\Omega ) = \sqrt{4\pi} Y_{lm} (\Omega )$, are proportional to
the spherical harmonics, $Y_{lm} (\Omega )$, which are indexed by
the integer pair $(l,m)$, with $l = 0,1, \ldots , + \infty $ and $ m
= - l, \ldots , + l$. Furthermore, the eigenvalues for the pure
chain model, $\lambda _l $, are indexed only by $l$, because they
are degenerate in $m$:
\begin{equation}
\lambda _l  = e^{ - \kappa } \kappa ^l \left( {\frac{1}{\kappa
}\frac{d}{{d\kappa }}} \right)^l \left[ {\frac{{\sinh (\kappa
)}}{\kappa }} \right]=\left( {\frac{\pi }{{2\kappa }}} \right)^{1/2}
e^{ - \kappa } I_{l + 1/2} (\kappa )
\end{equation}
where $ I_{l + 1/2} (\kappa )$ is the modified Bessel function of
the first kind. These eigenvalues take on the values $ \lambda _0 =
e^{ - \kappa } \sinh (\kappa )/\kappa = \exp[ - G_0 (\kappa )]$ and
$ \lambda _1 = \lambda _0 \,u(\kappa ) = \exp[ - G_1 (\kappa )]$ for
$l = 0,1$ and are related to the $l = 0,1$ 2-link free energies
already calculated, Eqs.~(\ref{G0}) and (\ref{G1})~\cite{joyce}.

The rotational symmetry of the coupled model implies that in this
case the eigenspinors can still be labeled by the indices
$(l,m;\tau)$ used for the pure Ising and pure chain models:
\begin{equation}
\langle \Omega | \hat \Psi_{l,m;\tau } \rangle  = \psi_{lm} (\Omega
)| l,\tau\rangle
\end{equation}
and the eigenvalues, $\lambda _{l,\tau } $ by $(l,\tau )$
(degenerate in $m$). The eigenvalues and  kets, $\left| {l,\tau }
\right\rangle$, which are independent of the solid angle, $\Omega $,
must be determined by solving the eigenvalue Eq.~(\ref{eigenvalpb}).
In general, the eigenvalues and eigenvectors for the coupled system
are not, however, simply direct products of the corresponding
eigenvalues and eigenfunctions of the uncoupled Ising and chain
systems. By solving the eigenvalue equation Eq.~(\ref{eigenvalpb}),
we find that the kets, $| l,\tau\rangle$, appearing in the full
eigenspinor, and the eigenvalues $\lambda_{l,\tau }$, have already
been introduced and obtained for $l = 0$ and 1
[Eqs.~(\ref{vpl})-(\ref{al})] in the calculation of the effective
Ising partition functions, Eq.~(\ref{Zl}). If we define $G_l (\kappa
) = - \ln \lambda _l$, then the same formul{\ae},
Eqs.~(\ref{vpl})--(\ref{al}), apply for the kets $ \left|{l,\tau }
\right\rangle$ and the eigenvalues $\lambda_{l,\tau}$ in the general
case $l = 0,1, \ldots , + \infty $, $\tau  =  \pm $. The
eigenspinors are orthonormal:
\begin{equation}
\langle\hat \Psi_{l,m;\tau }|\hat \Psi_{l',m';\tau '}\rangle =
\langle {l,\tau } | {l',\tau '} \rangle \int \frac{d\Omega}{4\pi}
\psi_{l'm'}^* (\Omega )\psi_{lm} (\Omega ) = \delta _{ll'} \delta
_{mm'} \delta _{\tau \tau '}.
\end{equation}

Once we have the eigenvalues and orthonormal eigenfunctions, we can
express the transfer kernel in an abstract operator notation as
\begin{equation}
\hat P = \sum_{l = 0}^{ + \infty }  \sum_{m =  - l}^{ + l}
\,\sum_{\tau  = \pm }  \,\lambda _{l,\tau }| { {\hat \Psi
}_{l,m;\tau } } \rangle \langle { {\hat \Psi }_{l,m;\tau } }|
\end{equation}
and then use the orthonormality of the eigenspinors, as well as the
decomposition of unity,
\begin{equation}
\hat I =  \sum_{\sigma} \int \frac{d\Omega}{4\pi} |\sigma
\Omega\rangle\langle \sigma \Omega|= \sum_{l = 0}^{ + \infty }
\sum_{m =  - l}^{ + l}  \,\sum_{\tau  = \pm }  \,| { {\hat \Psi
}_{l,m;\tau } } \rangle\langle { {\hat \Psi }_{l,m;\tau } }|
\end{equation}
to calculate the quantities of interest in a straightforward way. As
a check on the method, we can,  for example, recalculate the
partition function using the following expression:
\begin{equation}
\mathcal{Z} = (4\pi )^N \sum_{\{ \sigma _i  =  \pm 1\}} \,\prod_{i =
1}^N\,\int \frac{d\Omega _i}{4\pi} \langle {V|\sigma _1 \Omega _1 }
\rangle \langle {\sigma _1 \Omega _1 |\hat P|\sigma _2 \Omega _2 }
\rangle \cdots \langle {\sigma _{N - 1} \Omega _{N - 1} |\hat
P|\sigma _N \Omega _N }\rangle \langle {\sigma _N \Omega _N |V}
\rangle , \label{Zdevelop}
\end{equation}
or in kernel product form
\begin{equation}
\mathcal{Z}  =  (4\pi )^N \left\langle V \right|\hat P^{N - 1}
\left| V \right\rangle  = (4\pi )^N \sum_{l,m;\tau } \langle V |\hat
\Psi_{l,m;\tau }\rangle^2 \lambda _{l,\tau}^{N - 1}. \label{Zint}
\end{equation}

Because the end vector $| V \rangle $ contains only the rotational
ground state, $| \hat \Psi _{0,0; \pm }\rangle $ (i.e., $l = 0,m =
0$), its  matrix elements with the eigenspinors of the transfer
kernel simplify to
\begin{equation}
\langle { {\hat \Psi }_{l,m;\tau } }| V \rangle  = \delta _{l0}
\delta _{m0}\langle {0,\tau } | V \rangle. \label{eigenspinors}
\end{equation}
Inserting this expression for the matrix element into
Eq.~(\ref{Zint}) leads immediately to the result, Eq.~(\ref{Zl}),
obtained previously for  $l=0$:
\begin{equation}
\mathcal{Z} = \mathcal{Z}_{\rm I,eff}^{(0)} = (4\pi )^N \sum_{\tau =
\pm } \langle V | 0,\tau \rangle^2 \lambda _{0,\tau}^{N - 1}.
\label{partf}
\end{equation}
In order to calculate the correlation function $ \langle {\bf t}_i
\cdot {\bf t}_{i + r}\rangle$, we could use
\begin{equation}
{\bf t}_i  \cdot {\bf t}_{i + r}  = \frac13\sum_{m =  - 1}^{ + 1}
\psi _{1m}^* (\Omega_{i + r}) \psi _{1m} (\Omega _i )
\label{harmonics}
\end{equation}
which can be obtained from Eq.~(\ref{anglecomp}) and the definition
of the spherical harmonics. Thanks, however,  to rotational
symmetry, the average value of $ {\bf t}_i  \cdot {\bf t}_{i + r}$
simplifies to
\begin{equation}
\left\langle {{\bf{t}}_i  \cdot {\bf{t}}_{i + r} } \right\rangle  =
3\,\left\langle {t_{i,z}  \cdot t_{i + r,z} } \right\rangle  =
\left\langle {\psi _{10} (\Omega _i )\,\psi _{10} (\Omega _{i + r}
)} \right\rangle
\end{equation}
where $ \psi _{10} (\Omega ) = \sqrt{3} \cos( \theta)$. The
tangent-tangent correlation function can be written in a kernel
product form similar to the expression for the partition function,
Eq.~(\ref{Zdevelop}), with the difference being that we must now
insert the projection (or dipole) operator along the $z$-axis, $\hat
Z=\cos(\theta)$, related to $\psi_{10}(\Omega)$ in the $j = i$ and
$j = i + r$ positions. This operator, which is  diagonal in the
canonical basis, $|\sigma \,\Omega\rangle$,  has the following
matrix elements:
\begin{equation}
\langle {\sigma _i \Omega _i | \hat Z|\sigma _{i + 1} \Omega _{i +
1} }\rangle  =\frac1{\sqrt3} \psi _{10} (\Omega _i )\delta \left(
{\Omega _{i + 1}  - \Omega _i } \right)\delta _{\sigma _{i + 1}
\sigma _i}
\end{equation}
from which the following equality can be established
\begin{equation}
\psi _{10} (\Omega _i) \langle\sigma_i\Omega _i |\hat P
|\sigma_{i+1} \Omega_{i+1}\rangle = \sqrt3\sum_{\sigma} \int
\frac{d\Omega}{4\pi} \langle \sigma _i \Omega _i |\hat Z|\sigma
\Omega\rangle  \langle\sigma\Omega |\hat P |\sigma_{i+1}
\Omega_{i+1}\rangle = \sqrt3\langle\sigma_i\Omega _i |\hat Z \cdot
\hat P |\sigma_{i+1} \Omega_{i+1}\rangle.
\end{equation}
In operator product form, using
Eqs.~(\ref{Zdevelop},\ref{harmonics}) the correlation function then
becomes
\begin{equation}
\langle {\bf t}_i  \cdot {\bf t}_{i + r} \rangle  =  3(4\pi )^N
\mathcal{Z}^{-1} \langle V |\hat P^{i - 1}\,\hat Z\,\hat P^r \hat
Z\,\hat P^{N-r-i}| V \rangle. \label{tangentcorr}
\end{equation}
In the basis that diagonalizes the transfer kernel, $\hat P$ , the
operator $ \hat Z$, which is not diagonal, has the following matrix
elements:
\begin{equation}
\langle \hat \Psi_{l',m';\tau '} |\hat Z | \hat \Psi_{l,m;\tau }
\rangle = \langle {l',\tau '} | {l,\tau } \rangle \delta _{mm'}
\left[ \delta _{l',l - 1} \left( \frac{l^2  - m^2}{4l^2 -
1}\right)^{1/2}  + \delta _{l',l + 1} \left(\frac{(l + 1)^2  -
m^2}{4(l + 1)^2  - 1}\right)^{1/2} \right] \label{selectionrules}
\end{equation}
which, aside from the factor $ \langle {l',\tau '}| {l,\tau }
\rangle$, is the well known selection rule for quantum dipole
transitions, i.e., $\Delta l =  \pm 1$
 and $\Delta m = 0$ (in, for example, the Stark effect~\cite{stark}). Although
 $\langle {l,\tau '} | {l,\tau } \rangle  = \delta _{\tau \tau '}$ the matrix element
 $\langle l',\tau ' | l,\tau  \rangle$ is not necessarily zero for  $l \ne l'$ and
 $\tau  \ne \tau '$, because in this case the matrix element is between states of
 different rotational symmetry. This result indicates that the
measurement of the $z$-axis dipole moment can also induce a change
in the internal state, $\tau$, of the system.

Equation~(\ref{eigenspinors}) shows that in the expression for
$\langle{{\bf{t}}_i  \cdot {\bf{t}}_{i + r} } \rangle$, the $\hat Z$
projection operator can only connect a $(l = 0,m = 0)$ rotational
state with an $(l = 1,m = 0)$ one, or vice-versa (as in the Stark
effect for the 1s state of the hydrogen atom). To evaluate $\langle
{{\bf{t}}_i  \cdot {\bf{t}}_{i + r} } \rangle$ using
Eq.~(\ref{tangentcorr}) we therefore only need one matrix element,
\begin{equation}
\langle { {\hat \Psi }_{1,0;\tau '} |\hat Z| {\hat \Psi }_{0,0;\tau
} } \rangle  = \frac1{\sqrt3} \langle {1,\tau '} |{0,\tau}\rangle.
\end{equation}
By inserting the decomposition of unity between each matrix factor
in Eq.~(\ref{tangentcorr}) and using the orthonormality of the
eigenspinors, we obtain
\begin{equation}
\langle {\bf{t}}_i  \cdot {\bf{t}}_{i + r} \rangle  = \frac{(4\pi
)^N}{\mathcal{Z}} \sum_{\tau _1, \tau _2,\tau _3} \langle V | 0,\tau
_3 \rangle \lambda_{0,\tau _3 }^{i - 1} \langle 0,\tau _3 | 1,\tau
_2 \rangle \lambda _{1,\tau _2 }^r \langle 1,\tau _2 | 0,\tau _1
\rangle \lambda_{0,\tau _1}^{N - r - i} \langle 0,\tau _1| V
\rangle. \label{tangentcorrfinal}
\end{equation}
When $i = 1$ and $r = N - 1$, using again the decomposition of unity
in the $| l,\tau  \rangle $ space, we recover our previous result
for $ \langle{\bf t}_1 \cdot {\bf t}_N\rangle$,
Eq.~(\ref{tangentNcorr}).

The above expressions for $\langle {\bf t}_i  \cdot {\bf t}_{i +
r}\rangle$, Eqs.~(\ref{tangentcorr})--(\ref{tangentcorrfinal}) can
also be interpreted, using the ``path integral" representation of a
quantum statistical partition function, as a quantum mechanical
measurement process over an imaginary time period of $N$ steps. This
interpretation is based on the (\textit{1D classical Ising
representation}) $\otimes$ (\textit{1D classical Heisenberg})
representation of the partition function of a 0D quantum diatomic
molecule, modeled as a 2 state rigid rotator. The system is prepared
in an initial state $| V \rangle $ that is in a mixture of the
internal states, $\tau  =  \pm $, but in the spherically symmetric
rotational ground state. This initial state evolves $N - r - i$ time
steps under the dynamics determined by the propagator $\hat P$,
until a measurement of the dipole moment along the $z$-axis,
determined by the action of $\hat Z$, is performed. The projections
of $| V\rangle$ onto the rotational ground states $(l = 0,m = 0)$,
$\langle {0,\pm }| V \rangle$, evolve in a simple way because these
states are associated with eigenspinors of the transfer kernel (each
time step gives rise to an eigenvalue factor, $\lambda_{0,\tau }$).
This dipole measurement excites the system from the rotational
ground state to the $(l = 1,m = 0)$, non-spherically symmetric,
first rotational excited state, along with a possible transition in
the internal state of the diatomic molecule. The state that  emerges
from the measurement then evolves $r$ time steps, until a second
measurement of the dipole moment is performed. This measurement
de-excites the system from the $(l = 1,m = 0)$ excited state back
down to the $(l = 0,m = 0)$ ground state or up to the $(l = 2,m =
0)$ excited state  [see Eq.~(\ref{selectionrules})] again with a
possible change in internal state. The state that comes out of this
second measurement then evolves $i - 1$ further time steps.  From
the representation in Eq.~(\ref{tangentcorrfinal}) we see that the
correlation function $\langle {\bf{t}}_i  \cdot {\bf{t}}_{i + r}
\rangle$ is then the normalized amplitude that the system returns to
the initial state $| V\rangle $ at the end of this double dipole
measurement process [because the final state and the $(l = 2,m = 0)$
excited state are orthogonal, no $l=2$ matrix elements appear in
Eq.~(\ref{tangentcorrfinal})]. By resolving the initial and final
states, both equal to $| V\rangle $ , into $| 0, \pm \rangle$
components, the sum in Eq.~(\ref{tangentcorrfinal}) is over all
possible ``time" sequences involving $|0,\pm\rangle$.

In order to study the limiting forms of  $\langle{\bf{t}}_i  \cdot
{\bf{t}}_{i+ r}\rangle$, we write
\begin{equation}
\langle {{\bf{t}}_i  \cdot {\bf{t}}_{i + r} } \rangle  = \frac{
\sum_{\tau _1,\tau _2 ,\tau _3} C(\tau _3 ,\tau _2 ,\tau _1 )
\left(\frac{\lambda_{0,\tau _3}}{\lambda_{0,\tau _1 }}\right)^{i -
1} \left(\frac{\lambda_{1,\tau _2}}{\lambda_{0,\tau _1 }} \right)^r
\left(\frac{\lambda_{0,\tau _1}}{\lambda_{0, + }} \right) ^{N - 1} }
{   \sum_{\tau} \left(\frac{\lambda _{0,\tau}}{\lambda _{0,
+}}\right)^{N - 1} \langle V | 0,\tau\rangle^2},
\end{equation}
where we have introduced the joint amplitude
\begin{equation}
C(\tau _3 ,\tau _2 ,\tau _1 ) = \langle V | {0,\tau_3 } \rangle
\langle {0,\tau _3 } | {1,\tau_2 } \rangle \langle {1,\tau _2 } |
{0,\tau_1 } \rangle \langle {0,\tau _1 } | V \rangle.
\end{equation}
When all the bending rigidities are equal to $\kappa$, we recover,
as expected, the pure discrete wormlike chain result, $\langle
{{\bf{t}}_i  \cdot {\bf{t}}_{i + r} } \rangle  = \exp[- r/\xi _p
(\kappa)]$ with $\xi _p$  given in  Eq.~(\ref{xi}).

In the limit $N \to \infty $, we keep only the leading order term
$\propto \lambda _{0, +}^{N - 1} $ in $\mathcal{Z}$ and the
surviving terms in the numerator 2-point correlation function
($\tau_1=+$) to find:
\begin{equation}
\langle {\bf t}_i  \cdot {\bf{t}}_{i + r} \rangle
\underset{N\to\infty}{\to} \sum_{\tau _2,\tau _3} C'(\tau _3 ,\tau
_2 ) \left(\frac{\lambda_{0,\tau _3 }}{\lambda _{0, + }} \right)^{i
- 1} \left(\frac{\lambda _{1,\tau _2 }}{\lambda _{0, + }}\right)^r
\label{tangentcorrNinf}
\end{equation}
where
\begin{equation}
C'(\tau _3 ,\tau _2) = \frac{   \langle V | {0,\tau_3 } \rangle
\langle {0,\tau _3 } | {1,\tau _2 } \rangle \langle 1,\tau _2 |0,
+\rangle}{\langle0, + | V \rangle}.
\end{equation}

Using the effective chain persistence lengths introduced previously
in Eq.~(\ref{persists}), we can express Eq.~(\ref{tangentcorrNinf})
in a physically more transparent form:
\begin{equation}
\langle {\bf t}_i  \cdot {\bf t}_{i + r}\rangle
\underset{N\to\infty}{\to} \sum_{\tau _2 =\pm} \exp \left[- r/\xi
_{1,\tau_2 }^p \right] \left \{ C'(+,\tau _2 )+ C'(-,\tau _2 )\exp
\left[ - (i -1)/\xi_{I} \right] \right\} \label{tangentcorrNinf2}
\end{equation}
with $\xi _{1, - }^p  < \xi _{1, + }^p$  and $\xi _{I}$ the Ising
correlation length already introduced in Eq.~(\ref{ising}).

In the double limit $N,i \to \infty $, the dependence on the chain
ends disappears again and the expression
Eq.~(\ref{tangentcorrNinf2}) simplifies to
\begin{equation}
\langle {\bf{t}}_i  \cdot {\bf{t}}_{i + r}\rangle \underset{N,i
\to\infty}{\to} \sum_{\tau_2 = \pm} \langle 1,\tau_2 | 0,+ \rangle^2
\exp \left(- r/\xi_{1,\tau_2}^p\right), \label{tangentcorrNiinf}
\end{equation}
which reveals the importance of the two persistence lengths,
$\xi_{1, \pm}^p$, and the two ``transition probabilities",
$\langle1, \pm | 0, + \rangle^2$, for going from the ground state
$|0,+\rangle$ to the first rotational excited state,
$|{1,\pm}\rangle$, with or without a change in internal state
$\tau$. In the temperature range of experimental interest, $T <
T_1^\infty$, $\langle 1, + | 0, + \rangle^2$ and $\langle 1, - | 0,
+ \rangle^2$ can, to an excellent approximation, be set equal to
$\varphi_{U, \infty}$ and $\varphi_{B, \infty}$, respectively. When
this last result is used in conjunction with the limiting forms for
$\xi_{1,\pm}^p$, Eq.~(\ref{limpers}), a useful approximation is
obtained for Eq.~(\ref{tangentcorrNiinf}), valid for $T <
T_1^\infty$:
\begin{equation}
\langle {\bf{t}}_i  \cdot {\bf{t}}_{i + r}\rangle \underset{N,i
\to\infty}{\simeq} \varphi_{U, \infty} \exp \left(-
r/\xi_{1,+}^p\right) + \varphi_{B, \infty} \exp \left(-
r/\xi_{1,-}^p\right). \label{tangentcorrNiinfap}
\end{equation}

For $N,i \to \infty $ and short distances, $r \ll \xi _{1,-}^p$, we
find the limiting linear behavior in $r$:
\begin{equation}
\langle {\bf{t}}_i  \cdot {\bf{t}}_{i + r}\rangle \underset{N,i
\to\infty}{\simeq}
  1- r/\xi_{\rm eff, CF}^p,
\label{tangentcorrNiinflin}
\end{equation}
where
\begin{equation}
1/\xi_{\rm eff, CF}^p \equiv \langle 1,+ | 0,+ \rangle^2 / \xi
_{1,+}^p + \langle 1,- | 0,+ \rangle^2 / \xi _{1,-}^p \simeq
\varphi_{U, \infty} / \xi _{1,+}^p + \varphi_{B, \infty} / \xi
_{1,-}^p \label{effperscf}
\end{equation}
is an effective persistence length for the correlation function (CF)
at short distances that clearly  reveals the importance of the
shortest persistence length, $\xi _{1,-}^p$, under these conditions.
In the triple limit $N,i,r \to \infty $, only one term survives:
\begin{equation}
\langle {\bf{t}}_i  \cdot {\bf{t}}_{i + r} \rangle
\underset{N,i,r\to\infty}{\to} \langle {1, + } |0, + \rangle^2 \exp
\left(- r/\xi _{1, + }^p\right).
\end{equation}
Exactly at $T_m^\infty$, the above expression
Eq.~(\ref{tangentcorrNiinf}) simplifies to
\begin{equation}
\langle {\bf{t}}_i  \cdot {\bf{t}}_{i + r}\rangle_{T_m^\infty}
\underset{N,i \to\infty}{\simeq} \frac12 \left[ \exp
\left(-r/\xi_U^p\right) + \exp\left(-r/\xi_B^p\right)\right],
\label{tangentcorrNiinftm}
\end{equation}
which for $r \ll \xi_B^p$ reduces to
\begin{equation}
\langle {\bf{t}}_i  \cdot {\bf{t}}_{i + r}\rangle_{T_m^\infty}
\simeq 1-\frac{r}2 \left(1/\xi_U^p + 1/\xi_B^p \right).
\label{tangentcorrNiinftmlin}
\end{equation}
Because the inverse persistence lengths enter into
Eq.~(\ref{tangentcorrNiinflin}), the short distance limiting
behavior will tend to be dominated by the shortest one, $\xi _{1, -
}^p$ above $T_m^\infty$, where $\langle 1,- | 0,+ \rangle^2 >
\langle 1,+ | 0,+ \rangle^2$. Below $T_m^\infty$, however, there
will be a competition between the weights $\langle 1,- | 0,+
\rangle^2 < \langle 1,+ | 0,+ \rangle^2$ and the persistence
lengths, $1/\xi _{1, - }^p > 1/\xi _{1, + }^p$.

\begin{figure}[t]
\includegraphics[height=4.5cm]{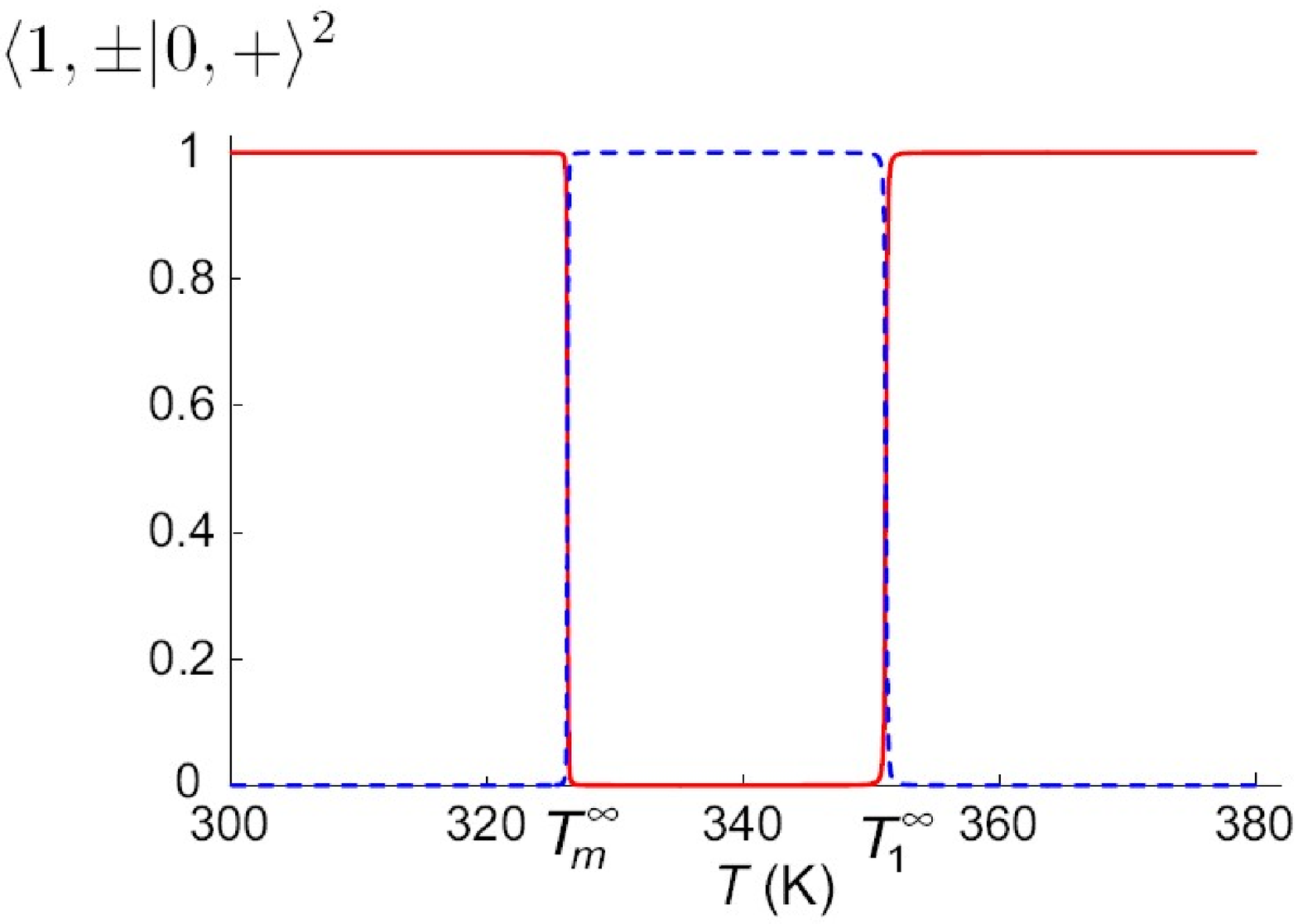}
\includegraphics[height=4.5cm]{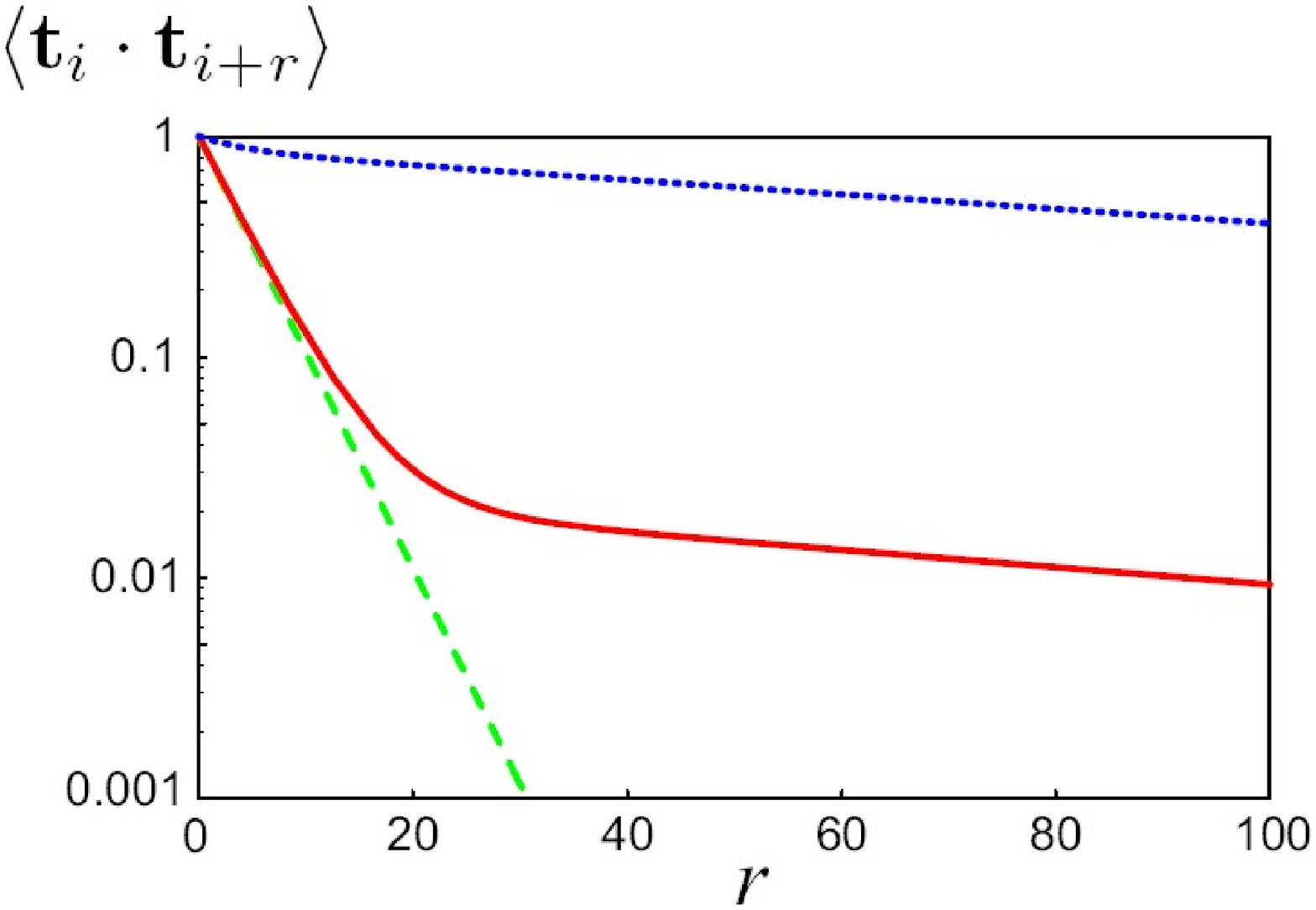}
\caption{(a)~Variation of the transition probabilities $\langle {1,
\pm} |0, + \rangle^2$ with temperature (red solid line for $+$ and
dashed blue line for $-$). For $T<T_1^\infty$, $\langle {1, +} |0, +
\rangle^2 \simeq \varphi_{U, \infty}$ and $\langle {1, -} |0, +
\rangle^2 \simeq \varphi_{B, \infty}$. One observes that for
$T_m^\infty<T<T_1^\infty$, $\langle1,-|0,+\rangle^2=1$ which
underlines the relevance of $\xi _{1, -}^p$ in this temperature
range (with parameter values $\tilde \mu=4.46$~kJ/mol, $\tilde
J=9.13$~kJ/mol and $\tilde K=0$, see section~\ref{DNA}).
(b)~Tangent-tangent correlation function given by
Eq.~(\ref{tangentcorrNiinf}) ($N,i\to\infty$) for 3 different
temperatures: just before the transition (dotted blue line),
controlled by $\xi_{1,+}^p  \simeq \left[ 1/\xi_{U}^p + 1/\xi_I
\right] ^{-1} \simeq \xi_{U}^p$; slightly above $T_m^\infty$ (solid
red line), where the correlation length $\xi _{1,-}^p \simeq
\xi_{B}^p$ is dominant for $r<r^* \simeq 20$; and after the
transition ($T_m^\infty<T<T_1^\infty$) where the correlation length
$\xi _{1,+}^p \simeq \xi_{I}$ disappears in favor of $\xi _{1,-}^p
\simeq \xi_{B}^p$ (dashed green line).} \label{fig3}
\end{figure}

The conditions under which these limiting expressions are valid
approximations depends critically on the weights  appearing in the
above expressions. If, for example, $ {\langle {1, + } | {0, + }
\rangle }^2 \simeq \varphi_{U, \infty}$ is sufficiently small
compared with $ {\langle {1, - } | {0, + } \rangle }^2 \simeq
\varphi_{B, \infty}$ at a certain temperature, then the
``subdominant" term in Eq.~(\ref{tangentcorrNiinf}), $\propto
\langle 1, - | 0, + \rangle^2$ (possessing the smaller persistence
length) may actually be dominant over a wide range of $r$ values, as
shown in Fig.~\ref{fig3}.

\begin{figure}[t]
\includegraphics[height=6cm]{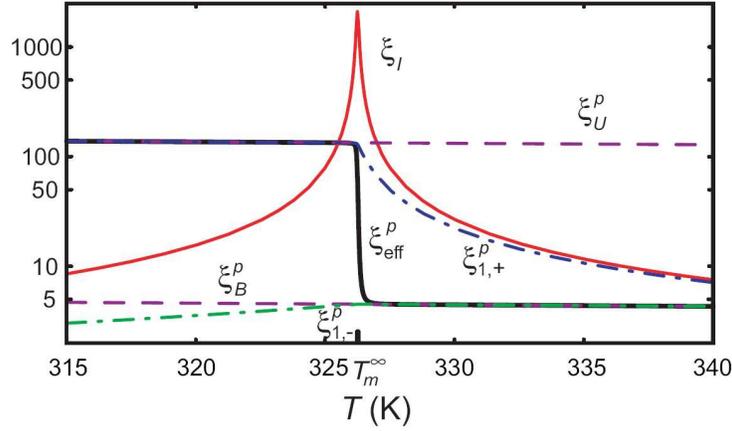}
\caption{Variation with temperature of the various correlation
lengths appearing in the model results:  the Ising correlation
length, $\xi_I$ (solid red line); persistence lengths of the coupled
system, $\xi^p_{\rm eff}$ (appearing for long chains) and
$\xi_{1,\pm}^p$ (dashed dotted line respectively blue and green);
and of the pure chains, $\xi^p_{U,B}$ (dashed purple lines) (in
units of $a$). At $T_m^{\infty}$, the Ising correlation length is
peaked but finite, which is related to the point of closest approach
of the two $(0,\pm)$ branches (zoom of Figure~\ref{fig2}), and the
effective persistence length, $\xi^p_{\rm eff}$, rapidly crosses
over from $\xi^p_U$ to $\xi^p_B$ (parameter values $\tilde
\mu=4.46$~kJ/mol, $\tilde J=9.13$~kJ/mol and $\tilde K=0$).}
\label{fig4}
\end{figure}

 Indeed, due to the coupling between bending and internal
states of DNA, for realistic parameter values (cf.
Section~\ref{DNA}), the respective weights $\lan 0,+|1,\pm\ran^2$
associated with each correlation length change abruptly at
$T_m^\infty$: below $T_m^\infty$, we have $\lan
0,+|1,+\ran\simeq\lan U|U\ran=1$ and $\lan0,+|1,-\ran\simeq\lan
U|B\ran=0$, thus $\xi_{1,+}^p\simeq \xi_U^p$. For  $T_m^\infty < T <
T_1^\infty$, we find $\lan 0,+|1,+\ran\simeq\lan B |U\ran=0$ and
$\lan 0,+|1,-\ran\simeq\lan B|B\ran=1$ which implies
$\xi_{1,-}^p\simeq \xi_B^p$. For higher temperatures,
$T>T_1^\infty$, the respective weights get swapped again, but now
$\xi_{1,+}^p\simeq \xi_{1,-}^p$. These considerations lead us to
introduce a critical distance, $r^*$, at which the two terms in
Eq.~(\ref{tangentcorrNiinf}) are equal:
\begin{equation}
r^* \equiv \left( \frac{1}{\xi _{1, - }^p} - \frac{1}{ \xi _{1, +
}^p } \right)^{-1} \ln \left(  \frac{\langle 1,+ | 0,+
\rangle^2}{\langle 1,- | 0,+ \rangle^2} \right) \simeq \xi _{B}^p
\ln \left( \frac{\varphi_{B, \infty}}{\varphi_{U, \infty}} \right),
\label{rstar}
\end{equation}
where in arriving at the last approximation we have used  limiting
forms that are valid when $T<T_1^\infty$ and assumed that $\xi
_{B}^p \ll \xi _{U}^p, \xi _{I}$ (cf. Fig.~\ref{fig4}). When $r <
r^*$ then the correlation function Eq.~(\ref{tangentcorrNiinf}) is
dominated by the shortest persistence length, $\xi _{1, - }^p$, and
when  $r > r^*$ the correlation function is dominated by the longest
one, $\xi _{1, + }^p$. For sufficiently long chains and temperatures
close enough to $T_m$, the inequality $N \gg r^*$ holds and this
cross over should be clearly visible (see Fig.~\ref{fig3}).

\section{Mean-Square end-to-end distance}
\label{Rg}

To calculate the mean-square end-to-end distance of the chain, we
use the two-point correlation function obtained above:
\begin{equation}
\left(\frac{R}{a}\right)^2  = \sum_{i,j= 1}^N \langle {{\bf{t}}_i
\cdot {\bf{t}}_j } \rangle = N + 2\sum_{i = 1}^{N - 1} \sum_{r =
1}^{N - i} \langle {\bf{t}}_i  \cdot {\bf{t}}_{i + r} \rangle.
\label{giration}
\end{equation}
When all the bending rigidities are equal to $\kappa $, we recover,
as expected,  the pure discrete wormlike chain result:
\begin{equation}
R^2(\kappa)  = a^2\,N\,W_N(u(\kappa))\,\,\mathrm{where}\,\, W_N (z)
= \frac{1 + z}{1 - z} -\frac{2z}{N}\frac{1 - z^N}{(1 - z)^2}
\end{equation}
with $u(\kappa)$  defined in Eq.(\ref{G1}). In the limit $N \to
\infty$,
\begin{equation}
R^2(\kappa) \underset{N\to\infty}{\to} a^2\,N\, \frac{1 + e^{ -
1/\xi _p (\kappa )} }{1 -e^{ - 1/\xi _p (\kappa )}}.
\end{equation}
More specifically, three distinct regimes can be identified:
\begin{equation}
R^2(\kappa )  \to \left\{ \begin{array}{*{20}c}
   a^2\,N,  \\
   2\,a^2\,N\,\xi _p (\kappa ),  \\
   a^2\,N^2,  \\
\end{array}
\begin{array}{*{20}c}
   \xi _p (\kappa ) \ll 1 \ll N  \\
   1 \ll \xi _p (\kappa ) \ll N  \\
   1 < N \ll \xi _p (\kappa)  \\
\end{array} \right.
\begin{array}{*{20}c}
   {\rm (freely\quad jointed \quad Gaussian)} \\
   {\rm (effective\quad  spin\quad  wave \quad Gaussian)}  \\
   {\rm (rigid)}.  \\
\end{array}
\end{equation}
Because $ \xi _p (\kappa ) = \xi _p (\beta \tilde \kappa )$ is
decreasing function of temperature, the pure DWLC will go from the
rigid to the effective Gaussian to the freely jointed Gaussian
regime as the temperature is raised.

For the coupled model the double summation in Eq.(\ref{giration})
can also be carried out and we find
\begin{equation}
\left( \frac{R}{a} \right)^2  = N + 2\frac{ \sum_{\tau _1,\tau _2
,\tau _3} C(\tau _3 ,\tau _2 ,\tau _1 )\,S_N (\lambda _{0,\tau _3
},\lambda _{1,\tau_2 }  , \lambda_{0,\tau _1 }) \left(
\frac{\lambda_{0,\tau _1 }}{\lambda _{0, + }} \right)^{N - 1} }{
\sum_{\tau} \left( \frac{\lambda _{0,\tau}}{\lambda _{0, +
}}\right)^{N - 1} \langle V | 0,\tau \rangle^2 }, \label{Rgeneral}
\end{equation}
where
\begin{equation}
S_N ( {x,y,z} ) = \left\{
\begin{array}{*{20}c}
N \frac{y}{x - y} - \frac{y}{x} \frac{ 1 - (y/x)^N}{ (1 - y/x)^2},\,\,\mathrm{for}\,\, x = z \\
z^{1 - N} \frac{T_N (y,z) - T_N (y,x)}{z - x}, \,\,\mathrm{for}\,\,x \ne z \\
\end{array} \right. ,
\end{equation}
with $T_N (x,y) = y^N \frac{ x/y - (x/y)^N }{1- x/y}$. In the limit
$ N \to \infty$ the above complicated expression for $R^2$
simplifies to an effective Gaussian form
\begin{equation}
R^2 \underset{N\to\infty}{\to} 2 a^2 N \xi^p_{\rm eff} \quad
\mathrm{where} \quad \xi^p_{\rm eff} \equiv \frac12\sum_\tau \langle
1,\tau | 0, + \rangle^2 \frac{1 + e^{-1/\xi_{1,\tau }^p}}
{1-e^{-1/\xi _{1,\tau }^p}} \label{Rsimple}
\end{equation}
is an effective "long chain" persistence length. This expression can
be also obtained by using the simplified $N,i \to \infty$ limiting
form for $\langle {{\bf{t}}_i  \cdot {\bf{t}}_{i+r} }\rangle$,
Eq.~(\ref{tangentcorrNiinf}), in the general formula for $(R/a)^2$,
Eq.~(\ref{giration}). It tends to a further limiting form when
$\xi_{1,\pm }^p \gg 1$:
\begin{equation}
 \xi^p_{\rm eff}  \to \sum_\tau \langle 1,\tau | 0, + \rangle^2 \xi
_{1,\tau }^p \simeq\left\{
\begin{array}{ll}
   \varphi_{U, \infty} \xi^p_U+\varphi_{B, \infty}
\xi^p_B,  &  T <  T_m^\infty \\
   \varphi_{U, \infty} \left[ 1/\xi^p_U + 1/\xi_I \right]^{-1} + \varphi_{B, \infty}
\xi^p_B,   &  T_m^\infty < T <  T_1^\infty  \\
\end{array} \right. \label{Rfinal}
\end{equation}
At $T_m^\infty$ this expression simplifies reduces to $\xi^p_{\rm
eff} \simeq \left(\xi_{U}^p + \xi_{B}^p \right)/2$ when $\xi_I \gg
\xi^p_U$, which is actually the case (Fig.~\ref{fig4}).

For $T  < T_m^\infty$, the longest persistence length dominates:
$\xi^p_{\rm eff} \simeq \xi_{1, + }^p \simeq \xi_{U}^p$. Above
$T_m^\infty$, however, we see once again that there may be a
competition between the persistence lengths, $\xi _{1, \pm }^p$, and
the ``transition probabilities", ${\langle {1,\pm  } | {0, + }
\rangle }^2$, appearing in Eq.~(\ref{Rfinal}). This competition
occurs now for $T > T_m^\infty$, contrary to what was found for the
short distance behavior of the 2-point correlation function,
Eq.~(\ref{tangentcorrNiinflin}), because the persistence lengths
themselves appear in Eq.~(\ref{Rfinal}), and not their inverses. For
$T_m^\infty<T<T_1^\infty$, depending on the weights, the smaller
persistence length, $ \xi _{1, - }^p$, may actually be dominant over
the larger one, $\xi _{1, + }^p$; if so, $\xi^p_{\rm eff} \simeq
\xi_{1, -}^p \simeq \xi_{B}^p$, which is actually the case when we
consider standard parameter values for dsDNA and ssDNA
(section~\ref{DNA}), as shown in Fig.~\ref{fig4}.

\section{Application to synthetic DNA thermal denaturation}
\label{DNA}

Melting or thermal denaturation profiles are experimentally obtained
by following the UV absorbance of a DNA solution while slowly
increasing the sample temperature. This method allows one to follow
the temperature evolution of the fraction of base-pairs that have
been disrupted, $\varphi_B(T)$. A typical profile has a sigmoid
shape possibly with bumps that could appear depending on  the DNA
sequence. Different Ising-type  models have been
proposed~\cite{wartmont,wartben,polscher,gotoh} for modeling
denaturation curves by focusing on the influence of the base pair
sequence, but they do not attempt to take into account properly the
fluctuations of the DNA chains themselves. Yet, chain fluctuations
increase with $T$ and play a crucial role in determining melting
profiles. Moreover, these fluctuations concern both stiff helical
segments and flexible coils with different bending rigidities.

In this section, we compare the model developed above, whose key
element is to account for internal state fluctuations on an equal
footing  with those of the chain, with a set of experimental data.
We focus on the evolution  of $\varphi_B(T)$ for a synthetic
homopolynucleotide polydA-polydT. Six independent parameters appear
in the theory: the polymerization index $N$, the three Ising
parameters $K$, $J$ and $\mu$ defined in Eq.~(\ref{H}) and
Fig.~\ref{fig1}, and bending moduli $\kappa_U$ for dsDNA  and
$\kappa_B$ for ssDNA. Note that we have also introduced a bending
rigidity $\kappa_{UB}$ for domain walls. However $\kappa_{UB}$
appears in the theory only in the effective cooperativity parameter
$J_0$. Thus changing $\kappa_{UB}$ is equivalent to varying the bare
$J$, i.e. the energetic penalty to create a wall. Without any lost
of generality, we choose to fix $\kappa_{UB}=\kappa_U$. Moreover, we
choose free boundary conditions for the end monomers, which is valid
unless their state is fixed by the  experimental
conditions~\cite{libch} (although any type of boundary conditions
can be treated using our model). Of the six parameters, three are
determined experimentally: $N$, $\kappa_U$, and $\kappa_B$.
Moreover, there is evidence  that stacking interactions in dsDNA and
ssDNA are of the same magnitude which justifies the choice of
$\tilde K=0$ adopted below~\cite{goddard}.

\begin{figure}[ht]
\includegraphics[height=6cm]{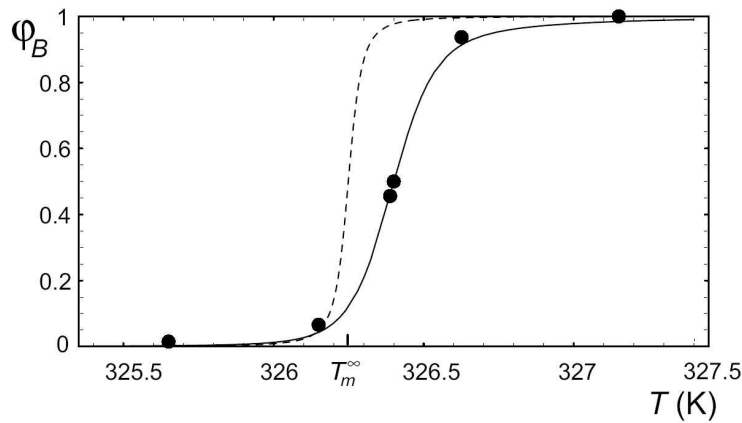}
\caption{Fraction of broken base-pairs for a polydA-polydT vs.
temperature (solution of 0.1 SSC, $N=1815$). The solid line
represents the theoretical law  for $\mu=1.64\,k_BT_m$ and
$J=3.35\,k_BT_m$ where $T_m=326.4$~K. The case $N\to\infty$ for the
same parameter values corresponds to the broken line.} \label{fig5}
\end{figure}

Figure~\ref{fig5} shows $\varphi_B(T)$ for a polydA-polydT of
molecular weight $M_w=1180$~kDa in a solution of 0.1~SSC (0.015 M
NaCl + 0.0015 M sodium citrate, pH 7.0) taken from~\cite{wartmont}.

In order to compare the data with our model predictions, we choose
the experimental values persistence lengths, $\ell_{ds}^p\simeq
50$~nm and $\ell_{ss}^p\simeq 1$~nm at 300~K, which lead to
$\kappa_U=\ell_{ds}^p/a=147$ and $\kappa_B=2\ell_{ss}^p/a=5.54$ at
$T=T_m$ (taking $a=0.34$~nm for one base-pair size and a factor of 2
for two  flexible segments in parallel per coil segment). The two
remaining parameters $\tilde \mu$ and $\tilde J$ are determined by
fitting the experimental data. The solid line in Fig.~\ref{fig5}
corresponds to $\tilde{\mu}=1.64\,k_BT_m\simeq4.46$~kJ/mol and
$\tilde{J}=3.35\,k_BT_m\simeq9.13$~kJ/mol leading to $T_m=326.4$~K.
We can then deduce several thermodynamical features. Noting that the
bare enthalpy for creating one A-T link is, in our model,
$2\tilde{\mu}$, we find a value very close to the experimental value
of 10.5 kJ/mol~\cite{pincet}. Although the value of $\tilde{J}$ is
more difficult to interpret, our result $\tilde{J}\sim 2
\tilde{\mu}$ is consistent with the idea that stacking interactions
make the dominant contribution to DNA stability~\cite{gotoh}. Chain
fluctuations do not only renormalize the effective free energy,
$2\tilde L_0$, required to break an interior base-pair, but also the
cooperativity parameter $\tilde J_0$: the latter varies almost
linearly with $T$ following Eq.~(\ref{J0}) contrary to previous
theories where $\tilde J$ was taken as constant and supposed to be
purely enthalpic in character~\cite{wartmont}. We have for the total
cooperativity parameter $\tilde{J}_0=4.17\,k_BT_m$ at $T=T_m$, which
shows that the bending contribution is roughly 25\% (remembering
that we have chosen $\kappa_{UB}=\kappa_U$). The model fit thus
leads to parameter values in accord with experiment. Our model
predictions for experimentally accessible A-T pair quantities are
also in agreement with accepted values~\cite{krueger,metzler2}: i)
the loop initiation, factor, $\sigma_{\rm LI} \equiv e^{-4J_0}
\simeq 10^{-7}$ at $T_m$;  and at physiological temperature, $T_{\rm
ph}$, ii)~ an interior single base-pair opening probability
$\varphi_B (T_{\rm ph}) \simeq 10^{-6}$ with a bubble initiation
barrier of $17 k_B T$ , and iii) a free energy of $0.18 k_B T$ for
breaking an additional base-pair in an already existing bubble. In
reality, the fitted values of $\tilde{\mu}$ and $\tilde{J}$
implicitly compensate for effects like loop entropy explicitly left
out of the model~\cite{wartmont}. As shown in Fig. 9 of
\cite{wartmont}, effective Ising models without loop entropy, like
ours, can be considered to account implicitly (and approximately)
for loop entropy, provided that one allows for loop entropy
contributions to both $J$ and $K$. This loop entropy renormalization
of the Ising model parameters will depend on the value of the loop
entropy exponent $k$ and the chain length $N$ and could allow for a
simple approximate  way of accounting for the influence of loop
entropy within the framework of an effective Ising model
(cf.~\cite{blossey}). This renormalization probably explains why our
the model value for $\sigma_{\rm LI}$ is at the low end of the
accepted spectrum.

In Fig.~\ref{fig5}, the curve corresponding to the thermodynamic
limit ($N\to \infty$) is shown for the same parameter values. In
this case, $\varphi_B(T)$ is given by Eq.~(\ref{c}) and the value of
the melting temperature is obtained analytically as a function of
$T_m^\infty$ using $L_0(T_m^\infty)=0$ which is given in the limit
of low temperature ($\tilde \kappa_B \gg k_BT$) by
\begin{equation}
k_BT_m^\infty\simeq 2\frac{\tilde{\mu}+\tilde{K}}{\ln(\kt_U/\kt_B)}
\label{Tminf}
\end{equation}
Hence, the melting temperature is reached when the enthalpy required
to create a link is perfectly balanced by the difference in (entropy
dominated) free energy between  the two types of semi-flexible
chains (U or B). Another quantity which has an experimental
relevance is the width of the transition. In the thermodynamic limit
this width is narrow, but nonzero, due to the large but finite
cooperativity parameter: $\Delta T_m^\infty \propto 1/\xi_I
(T_m^\infty) \simeq 2\exp[-2 J_0(T_m^\infty)]$ [see
Eq.~(\ref{deltaTm})]. Hence, the thermodynamic limit clarifies the
role of the two free model parameters: in conjunction with the
experimentally known bending rigidities, $\mu$ sets the melting
temperature and $J$ fixes the transition width. This is in contrast
to previous Ising-like models, where three fitting parameters were
used $J$, $\partial L_0/\partial T$, and $T_m^\infty$ with $L_0$
assumed to by a linear function of $T$~\cite{wartmont}.

Within the scope of our model the measured transition width is
indicative of   a very long Ising correlation length, $\xi _I$, near
the transition temperature, much larger than the pure U and B
persistence lengths;   therefore typical helix (U) and bubble (B)
domains (of size $\sim\xi _I$)  are flexible  within a small
temperature window near the transition.

Included in the predictions of our theory are mechanical and
structural features of the composed chain, such as persistence
length or mean square end-to-end radius, $R$. This differs from
purely Ising-type models~\cite{wartmont,gotoh} and non-linear
microscopic models~\cite{peyrard,dauxois} where only thermodynamical
quantities related to base-pairing are available. The variation of
the effective persistence length $\xi^p_{\rm eff}$ (and thus the
radius of gyration for long chains) vs. $T$ is shown in
Fig.~\ref{fig4}. It varies from $\xi^p_U$ for $T<T_m^\infty$ to
$\xi^p_B$ for $T>T_m^\infty$. Since the transition is very abrupt,
we suggest that the denaturation transition can also be followed
experimentally by measuring directly the radius of gyration, for
instance by tethered particle motion~\cite{pouget}, light
scattering, or viscosity experiments. For instance, since the
\textit{relative viscosity} is proportional to $c_{\rm DNA} R^3$
(where $c_{\rm DNA}$ is the DNA concentration), it should clearly
exhibit an abrupt thermal transition at a given $c_{\rm DNA}$ and
$N$. Such a transition has indeed been observed for the viscosity of
synthetic homopolynucleotide solutions~\cite{inman}, in qualitative
agreement with Fig.~\ref{fig4}.

In fitting our model to experiment for chains of length $N=1815$, we
have found that finite size effects play an important role
(Fig.~\ref{fig5}). In the following section, we investigate such
effects in detail.

\section{Finite size effects}
\label{finite:size}

\begin{figure}[t]
\includegraphics[height=6cm]{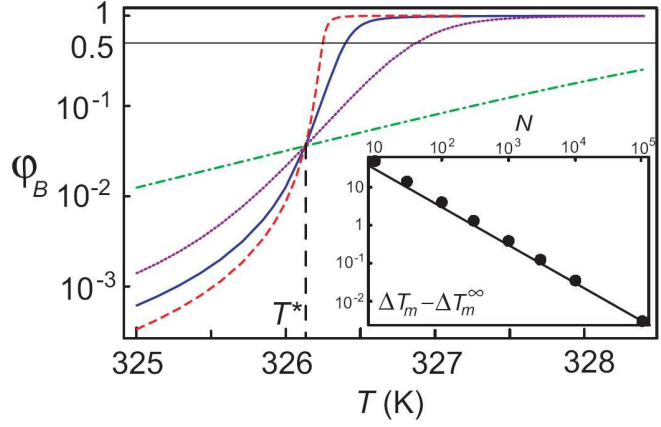}
\caption{Linear-Log plot of melting curves for $N= 100$ (dashed
dotted green line), 500 (dotted purple line), 1815 (solid blue
line), $\infty$ (dashed red line) in decreasing order at low
temperature, $T < 326$~K (parameter values $\tilde \mu=4.46$~kJ/mol,
$\tilde J=9.13$~kJ/mol and $\tilde K=0$). We note that all the
curves intersect at $T^*$, as discussed in the text. $T_m$ is
defined by $\varphi_B=0.5$. Inset: Log-Log plot of model results for
the shift in transition width $\Delta T_m-\Delta T_m^{\infty}$ vs.
polymer length. Dots correspond to the model results and the solid
line is a law in $1/N$.} \label{fig6}
\end{figure}

It has been shown experimentally that DNA thermal denaturation
varies with chain length, $N$~\cite{blake}. In this section, we
carefully study the effect of chain ends on the denaturation
transition. In Fig.~\ref{fig6} are shown denaturation profiles for
various chain lengths from $N=100$ to $N\to \infty$ and the fixed
parameters values $\tilde \mu=4.46$~kJ/mol, $\tilde J=9.13$~kJ/mol
and $\tilde K=0$ used in the previous section to fit the melting
data. Within the scope of our model one observes that i) the melting
temperature $T_m(N)$ is a decreasing function of $N$, varying as
$[T_m(N)- T_m^{\infty} ]/T_m^{\infty} \simeq 1/(N-1)$; ii) all the
denaturation curves intersect at a temperature $T^*$ at which
$\varphi_B \simeq 0.03$; iii) the transition width $\Delta T_m(N)$
is a decreasing function of $N$.

Concerning points (i) and (ii), the observed behavior for the
coupled system with the present parameter values is  directly
related to the model result that $T^* < T_m^\infty$, which is
radically different from the behavior found for the simple Ising
model (for which melting curves, $\varphi _{B}$, are strictly
decreasing functions of $N$, because formally, $T^* = \infty$ when
$\kappa _{U}=\kappa _{B}=\kappa _{UB}$). The present behavior for
melting maps is also very different from the predictions of older
empirical Ising-like models of denaturation and Helix-Coil like
transitions~\cite{nelson}, for which the chemical potential $\mu$
appearing in the end vector $|V \rangle$ is incorrectly identified
with $L_0$. This identification results in  melting curves
independent of $N$, i.e., $T_m= T^*$.

Concerning point (iii), the transition width roughly follows the law
$(\Delta T_m-\Delta T_m^{\infty})\sim 1/(N-1)$, which is a classical
result for finite size systems where fluctuations decrease in the
thermodynamic limit. One observes that even for a long polymer,
$N\sim 10^3$, finite size effects are important. For very short
chains, e.g., $N < 100$, such effects get amplified and we predict a
transition width as large as $50$~K for $N = 10$. This point is
crucial, since it has been observed experimentally that for
polydA-polydT inserts between more stable G-C rich domains, melting
curves are much wider for very short DNA chains ($N \sim 10$
bp)~\cite{libch}  with a width that decreases with increasing $N$
(observed for $60 < N < 140$ in~\cite{blake}). In such experiments,
the nature of end monomers clearly becomes extremely important.

\begin{figure}[t]
\includegraphics[height=6cm]{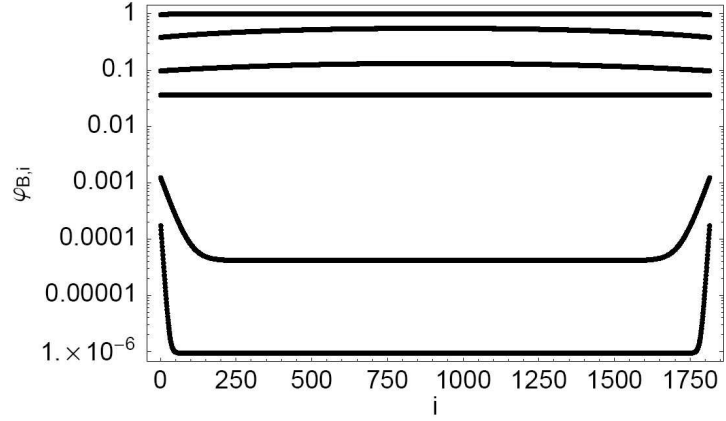}
\caption{Average melting maps for different temperatures for
$N=1815$ and the same parameter values used in Fig.~\ref{fig5}: plot
of the fraction of broken bases, $\varphi_{B,i}$ as a function of
the base position $i$ for, in increasing order, 310~K, $0.99\, T^* =
322.87$~K, $T^*=326.13$~K, $T_m^\infty=326.24$~K, $T_m=326.4$~K and
$1.01\,T_m = 329.66$~K.} \label{fig7}
\end{figure}

For a given $N$ and $T$ the local site dependent bubble opening
probability, or melting map,
\begin{equation}
\varphi_{B,i} = \frac{1 -\langle \sigma_i \rangle}2, \label{phib}
\end{equation}
can be obtained from $\langle \sigma _i \rangle$ given in
Eq.~(\ref{isingave}). In Fig.~\ref{fig7} $\varphi_{B,i}$ is plotted
for six different temperatures using the same model parameter values
employed in Fig.~\ref{fig5}; we observe that below $T^*$ the chain
unwinds from the ends, whereas above this temperature an interior
bond has a higher probability of being open than an end one. Far
enough below  $T^*$ the melting curve heals rapidly to a plateau
value close to $\varphi_{B,\infty}$ on a length scale on the order
of $\xi_I \ll N$. At physiological temperature, $310$~K, the
interior bond opening probability is $\simeq 10^{-6}$ in agreement
with the experimental value  for long runs of A-T pairs (which is an
order of magnitude lower than randomly placed A-T
pairs)~\cite{krueger}. At this temperature the end bonds have
opening probabilities two orders of magnitude greater than the
interior ones.

At $T^*$ the melting curve is perfectly flat. This result, which  is
independent of chain length $N$,  indicates that each Ising variable
can be considered to fluctuate independently: $\langle \sigma _i
\rangle$ is constant independent of $i$, despite a 2-point
correlation function that does not factorize [$\langle \sigma_i
\sigma_{i+r}\rangle \neq \langle \sigma_i \rangle \langle
\sigma_{i+r}\rangle$, cf. Eq.~(\ref{isingcorrsimple})] and a large
Ising correlation length ($\xi_I\gg1$, see Fig.~\ref{fig4}). Indeed,
the influence of the renormalized stacking energy ($\sim \tilde{K}_0
< 0$), which favors bubble formation, exactly compensates that due
to the renormalized destacking ($\sim \tilde{J}_0$), which
suppresses bubble formation, and therefore $\varphi _{B,i} (T^*) =
[1 - \tanh(\tilde{\mu}/(k_B T^*) ]/2 \simeq 0.03$, which results
from taking $N =1$ or taking $ K_0 = J_0 = 0$ for arbitrary $N$. In
some ways this compensation leads to an effective non-interacting
Ising system with $( \partial \langle c \rangle /
\partial N )_{T^*} = ( \partial \varphi _{B} / \partial N )_{T^*} =
0$, which  explains why the melting curves for different values of
$N$ cross at $T^*$ in Fig.~\ref{fig6}. Below $T^*$ the combined
effects of the renormalized stacking energy and entropy gain
favoring interior bubbles  are too small  to overcome the destacking
energy cost associated with an extra domain wall and the chain ends
unwind first. Since $K_0$ becomes more negative with increasing $T$
faster than $J_0$ increases, a temperature $T^*$ is reached where
the renormalized stacking and destacking effects just compensate.
Above $T^*$ the situation is reversed and the opening probability is
higher in the chain interior. For arbitrary $N$ and $T$, the
thermodynamic chemical potential, defined by $ \hat{\mu} = (\partial
F/\partial N)_T$ can be related to $\langle c \rangle $ \textit{via}
Maxwell-type relations, leading to $(\partial \hat{\mu}/\partial
\tilde{\mu} )_{T, N} = N (\partial \langle c \rangle / \partial N)_T
- \langle c \rangle$. At $T^*$ this general relation simplifies to $
(\partial \hat{\mu}/\partial \tilde{\mu} )_{T^*, N} = - \langle c
\rangle|_{T^*} = (\partial f / \partial \tilde{\mu} )_{T^*, N}$,
characteristic of a non-interacting system.

Upon examination of Eqs.~(\ref{isingave}) and (\ref{partf}), we see
that  $T^*$ is determined by the condition that $\left\langle
{\sigma _i } \right\rangle = \langle c \rangle_\infty$, which is
obtained when the end vector $| V \rangle$ is identical to the
eigenket $| 0,+ \rangle$ and orthogonal  to $| 0,- \rangle$:
$\langle V |0,+ \rangle = 1$ and $\langle V |0,- \rangle = 0$.
Physically, this means that the coupled Ising-chain system is in a
pure state, $| 0,+ \rangle$,  that mixes the canonical states in a
special way. The temperature $T^*$ can be obtained by solving
$\langle V |0,- \rangle = 0$. Using  Eqs.~(\ref{endv})
and~(\ref{vectpl})-(\ref{al})  this translates into $e^\mu =  e^{ -
2J_0} \{ \sinh(L_0 ) + [\sinh^2 (L_0 ) + e^{ - 4J_0}]^{1/2}\}$.
After some manipulation using Eq.~(\ref{c}), this can be shown to be
identical to $\langle c \rangle_\infty (T^*)= \langle c \rangle
(N=1, T^*) =\tanh[\tilde{\mu}/(k_B T^*)]$. Furthermore,
Eqs.~(\ref{isingcorr}) and (\ref{tangentcorrfinal}) show that the
Ising and chain 2-point correlation functions, $\langle {\sigma}_i
\cdot {\sigma}_{i+r}\rangle$ and $\langle {\bf t}_i \cdot {\bf
t}_{i+r}\rangle$, also get simplified at $T^*$: the approximate
forms, Eqs.~(\ref{isingcorrsimple}) and (\ref{tangentcorrNiinf})
valid in general only for $N, i \rightarrow \infty$, become exact
for arbitrary $N$ and $i$ at this special temperature. It is clear
that at $T^*$ the coupled system behaves as if there are no end
effects and finite chains have the same behavior as an infinite one.

\begin{figure}[t]
\includegraphics[height=5.5cm]{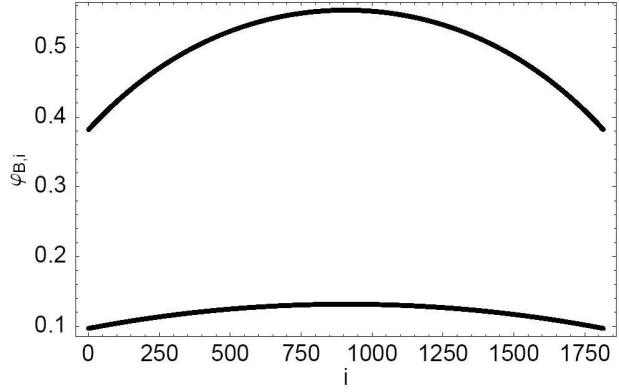}
\caption{Zoom of the melting map shown in Fig.~\ref{fig7} for
$T_m^\infty$ (lower curve) and $T_m$ (upper curve).} \label{fig8}
\end{figure}

To shed additional light on this mechanism and illustrate the
important role of internal bubble entropy for long chains, we now
study an infinite chain and compare $\varphi_{B, {\rm int}} =
\lim_{N,i \rightarrow \infty} \varphi _{B, i}$ with $\varphi_{B,
{\rm end}} = \lim_{N \rightarrow \infty} \varphi _{B, 1}$.
Equation~(\ref{isingaveinf}) shows that $\xi_I$ plays here the role
of a {\em healing length}, over which end effects relax (see
Fig.~\ref{fig7}). The ratio of matrix elements appearing in
Eq.~(\ref{isingaveinf}) gets simplified in the following way for
special values of $T$:
\begin{equation}
R_V = \frac{ \langle V |0,- \rangle }{ \langle V |0,+ \rangle } =
\left\{
\begin{array}{ll}
   -e^{-\mu},  &  T <  T^*  \\
   0,   &  T =  T^*  \\
   \tanh(\mu/2),  &  T =  T_m^\infty  \\
   e^{\mu},  &  T >  T_m^\infty \\
\end{array} \right.\label{specialv}
\end{equation}
At $T_m^\infty$, Eq.~(\ref{isingaveinf}) simplifies to
\begin{equation}
\left\langle {\sigma _i } \right\rangle_{\infty} (T_m^\infty) =
\tanh\left(\frac{\tilde{\mu}}{2 k_B T_m^\infty}\right)  \exp [
-(i-1)/\xi_I (T_m^\infty)],
\end{equation}
which shows that for very long chains ($N \gg \xi_{I}$) at
$T_m^\infty (> T^*)$ $\varphi _{B,{\rm end}} =
[1-\tanh[\tilde{\mu}/(2 k_B T_m^\infty)]/2 \simeq 0.16 \ll \varphi
_{B, \infty} (T_m^\infty) = 1/2$, revealing an internal opening
probability more than three times higher than an end one. In
Figs.~\ref{fig7} and \ref{fig8}, however, we observe that for $T =
T_m$ and $T_m^\infty$, $\xi_I>N = 1815$ (cf. Fig.~\ref{fig4}), and
therefore end effects do not get damped out near the center of this
finite chain. Indeed, at $T_m^\infty$ the opening probability
$\varphi_{B, i}$ near the middle of a chain of length $N = 1815$ is
much less than the value of 1/2 holding for an infinite one. For $N
= 1815$ we still observe, however, noticeable differences ($\sim 10$
to 20\%) between interior and end opening probabilities near $T_m$.

The \textit{one-sequence-approximation} has been defined by Poland
and Scheraga~\cite{polscher} and consists in neglecting the many
small bubbles which eventually collapse and considering only one
large thermally excited bubble. It is valid for temperatures
sufficiently far below $T_{m}^{\infty}$. For $N \to \infty$,
$\varphi_{B, {\rm int}}$ and $\varphi_{B, {\rm end}}$ can be
estimated in this approximation by summing over, respectively, all
the interior and end bubbles containing the fixed site in question.
In the interior case we thus have
\begin{equation}
\varphi _{B, {\rm int}} \simeq \sum_{n = 1}^{\infty} n \exp \left[
-\beta \Delta G_{\rm int}^{(n)} \right]= e^{-4J_0}\sum_{n =
1}^{\infty} n\; e^{-2nL_0}= \frac{e^{-4J_0}}{4
\sinh^2(L_0)}.\label{phiint}
\end{equation}
The factor of $n$ in the sum is entropic in nature and equal to the
number of ways of placing a fixed interior site within an interior
$n$-bubble.  In the end case, where there is no entropic factor,
\begin{equation}
\varphi_{B,{\rm end}} \simeq \sum_{n = 1}^{\infty} \exp \left[
-\beta \Delta G_{\rm end}^{(n)} \right] =e^{K_0-2J_0}\sum_{n =
1}^{\infty} e^{-2nL_0} =\frac{ e^{-(2 J_0+\mu)}}{2 \sinh(L_0)}.
\label{phiend}
\end{equation}
It is important to note that for the model parameters employed, the
additional dimensionless free energy for breaking an additional
base-pair in an already existing bubble, $2L_0\simeq 0.18$ at
physiological temperature, is less than one and much smaller than
the bubble initiation energy cost for an end bubble, $\sim 2 J_0
\simeq 8$ (and {\em a fortiori} for an interior one, $\sim 4 J_0$).
This implies that even at physiological temperature, where the
probability of bond opening is very small, bubbles covering a wide
range of sizes contribute to the sums in
Eqs.~(\ref{phiint})-(\ref{phiend}): both $\langle n \rangle _{\rm
int} = \coth(L_0) \simeq  1/L_0 \simeq 6$ and $\langle n
\rangle_{\rm end}  = e^{-L_0}/[2 \sinh(L_0)] \simeq  1/(2|L_0|)
\simeq 3$ are of the same order of magnitude as $\xi_I$, which
varies as $1/(2|L_0|)$ when $|L_0| \ll 1$. Because $L_0$ decreases
with $T$ (going to 0 at $T_m^\infty$), both $\varphi _{B, {\rm
int}}$ and $\varphi _{B, {\rm end}}$ are increasing functions of
temperature. For $T < T^*$ the bubble initiation energy cost
dominates and $\varphi _{B, {\rm int}} < \varphi _{B, {\rm end}}$.
Thanks to the entropic factor, however, $\varphi _{B, {\rm int}}$
increases more rapidly than $\varphi _{B,{\rm end}}$ and the two
curves cross over at $T^*$, an estimate of which can be obtained by
equating the above two one-sequence-approximations for both these
quantities. As a check on the preceding discussion, the
one-sequence-approximations for infinite chains obtained above for
end and interior opening probabilities can be shown to be  in exact
agreement with what one gets from the definition of $\varphi _{B,
i}$, Eq.~(\ref{phib}), and Eq.~(\ref{isingaveinf}) by using the low
temperature approximations for $\langle V |0,- \rangle / \langle V
|0,+ \rangle \simeq - e^{-\mu}$ [Eq.~(\ref{specialv})] and $\langle
c \rangle_\infty$, Eq.~(\ref{c}), i.e. expanding to lowest order in
$e^{-4J_0}$, assuming that $e^{-4J_0} \ll \sinh^2(L_0)$, which is
the formal criterion for the validity of the
one-sequence-approximation.

Using the one-sequence-approximation, it is now easy to see how the
cost in loop entropy associated with internal bubbles will modify
the above results. In the so-called loop entropy
models~\cite{wartmont,polscher,polscher}  the internal bubbles
formed by the single strands are visualized as one polymer loop,
whose entropic cost has been estimated first by Zimm~\cite{zimm}.
Hence, although $\varphi_{B,{\rm end}}$ does not change, the
interior opening probability does, becoming
\begin{equation}
\varphi_{B, {\rm int}}^{\rm LE} \simeq \sum_{n = 1}^{\infty} n/(n_0
+ n)^{k} \exp [-\beta \Delta G_{\rm int}^{(n)}],
\end{equation}
where we have adopted a common simplified form for the loop entropy
factor, $f_{\rm LE}(n) = (n_0 + n)^{-k}$ parametrized by a constant
$n_0$, which may be as large as 100 \cite{blake}, and an exponent
$k$, usually assumed to be in the range $3/2 \leq k \leq 2.1$,
depending on the extent to which chain self-avoidance is taken into
account~\cite{peliti}. A  large value for $n_0$ would severely
reduce the importance of loop entropy for short chains and probably
reflects the presence of strong bending rigidity effect (an
important open question concerns  how to incorporate bending
rigidity into $f_{\rm LE}$ in a physically correct way).  Loop
entropy clearly lowers the probability of interior bubble opening
and will lead to an increase in  $T^*$. The situation is further
complicated if strand sliding, which may be important for periodic
DNA, is taken into account. For homopolymeric DNA like
polydA-polydT, strand sliding leads to a modified loop entropy
exponent, $k' = k - 1$~\cite{wartmont, polscher, orland2}, resulting
in a significant decrease in the importance of loop entropy. By the
way, it also minimizes the importance for homopolymeric DNA of
recent claims that a true first-order phase transition should occur
for infinite chains because $k$ appears to be greater than 2 when
self-avoidance is fully taken into account~\cite{peliti}. The
combined effects of loop entropy and strand sliding will lead to
increases in both $T^*$ and $T_m^\infty$. Neither the theoretical
nor the experimental situation concerning $T_m(N)$ is entirely clear
for homopolymeric DNA with free ends and further careful experiments
are clearly called for. If loop entropy and strand sliding are
included in the coupled Ising-chain model, $T^*$ might become higher
than $T_m^\infty$, which would imply that $T_m(N)$ would increase
with $N$~\cite{polscher}, unlike what what we find to occur when
these two effects are neglected.

In the future, we intend to examine these questions by incorporating
loop entropy and strand sliding  directly into our DNA model. With
all other factors being equal, adding loop entropy and strand
sliding increases the stability of the closed state, resulting in
sharper melting curves and higher values of $T_m(N)$ (see Figs.~9
and~10 of~\cite{wartmont} and \cite{blossey}). The importance of
this loop entropy contribution would change with chain length $N$
and may lead to an increase in $T_m(N)$ with increasing
$N$~\cite{polscher} at least for a certain range of chain sizes .

\section{Summary and Conclusion}
\label{cl}

This paper presents a novel theoretical model of DNA denaturation,
already introduced in~\cite{prl}, which focuses on the coupling
between the base-pair link state (unbroken  or broken) and the
rotational degrees of freedom of the semi-flexible chain.  The
Hamiltonian includes local chain bending rigidities whose values
depend on neighboring base-pair states: around 5~$k_BT$ for bubbles
and 150~$k_BT$ for connected base-pair segments. Because of the
rotational symmetry, the model can be rewritten in terms of an
effective Ising Hamiltonian by integrating out the rotational
degrees of freedom of the chain. Hence, our model yields
considerable insight into the empirical temperature-dependent
parameters used in previous Ising-like models~\cite{wartmont}. In
particular, the melting temperature $T_m$ is no longer a fitting
parameter, but emerges naturally as a function of: (i)
experimentally known bending rigidities, $\kappa_U$ and $\kappa_B$;
(ii) the bare energy required to open a base-pair, $2\tilde \mu$;
(iii) the bare energy of a domain wall, or destacking, $2\tilde J$;
(iv) the difference in bare stacking energy between ss and dsDNA,
$2\tilde K$; and (v)  the polymerization index, $N$. Moreover, our
model allows structural features of the DNA chain, such as the mean
size $R$, to be calculated as a function of $T$. An abrupt
transition for $R$ is found at $T_m$ and explains, at least
qualitatively, the thermal transition observed in viscosity
measurements~\cite{inman}.

From an experimental perspective, our results obtained from exactly
solving the coupled model can be summarized as follows. First of
all, we propose  formul{\ae} for chain free-boundary conditions,
this information being encoded in the end vector $|V \rangle$ given
in Eq.~(\ref{endv}). However, any other boundary condition can be
treated following the same route, even though we shall not detail
the calculations here. For example, a polydA-polydT sequence of
length $N$ sandwiched between more stable G-C sequences~\cite{libch}
can be seen near its melting transition as a DNA of length $N$, with
fixed boundary conditions, resulting in an end vector $|V \rangle =
|U \rangle$.

Once boundary conditions are set, our model predicts melting
profiles, as measured for example from UV absorbance experiments.
Even if the model relies upon six microscopic parameters, as
discussed in section VII, most of them are known experimentally,
including the strand length $N$, and only two of them must be
extracted from melting profiles: $\tilde \mu$, the bare half-energy
required to break a base pair (which can also be estimated
experimentally \cite{pincet}), and $\tilde J$, the cooperativity
parameter that indicates the cost of creating a domain wall between
unbroken and broken base pairs (recent experiments on single DNA
molecules aim at determining this quantity, see~\cite{ke}). Melting
profiles, giving the fraction of broken base pairs, $\varphi_B$, as
a function of the temperature $T$, are determined through the
average of the Ising state variable, $\langle c \rangle$, because
$\varphi_B (N, T) = [1-\langle c \rangle (N, T)]/2$.

For \textit{infinite chains}, ($N \rightarrow \infty$), the
expression for $\langle c \rangle_{\infty} (T)$ is rather simple
(Eq.~(\ref{c})):
\begin{equation}
\left\langle c \right\rangle_\infty (T) =
\frac{\sinh(L_0)}{[\sinh2(L_0) + e^{-4J_0}]^{1/2}},
\end{equation}
where the renormalized parameters  $J_0$,  $K_0$, and $L_0 =\mu +
K_0$ are given in Eqs. (\ref{J0}), (\ref{K0}), and (\ref{L0}). In
these latter equations, the function $G_0(\kappa)$ has a simple
algebraic expression (Eq.~(\ref{G0})) that reduces to a purely
entropic contribution, $G_0(\kappa) \simeq \ln(2\kappa)$, in the
physically relevant low-$T$ (spin wave) approximation. From the
melting profile $\varphi_{B, \infty} (T)$, the melting temperature,
$T_m^\infty$, is defined by $\varphi_{B, \infty} = 1/2$, in other
words by $L_0=0$. The finite transition width is estimated by
Eq.~(\ref{deltaTm}): $\Delta T_m^{\infty} \simeq 2\,k_B
[T_m^{\infty}]^2 \exp [-2\,J_0(T_m^{\infty})] /{\tilde \mu} $.

For \textit{finite length strands} ($N$ finite), the expression for
$\langle c \rangle (N, T)$, Eq.~(\ref{cn}), is more complex, because
$N$ has an influence on $\varphi_B (N, T)$ and thus $T_m$. In
addition, the interplay with mechanisms, such as loop entropy, not
taken into account in this study is nontrivial, as discussed in
detail in section~\ref{finite:size}. At the level tackled in the
present paper, we obtain a simplified expression for $\langle c
\rangle (N, T)$ when  $N$ is large, Eq.~(\ref{cninf}), which
simplifies even further when $N \gg \xi_I \gg 1$ (Fig.~\ref{fig7}):
\begin{equation}
\langle c \rangle (N,T)  \simeq \langle c \rangle_{\infty}  + 2 R_V
(\xi_I/N) \;  ( 1 - \langle c \rangle_\infty^2)^{1/2},
\end{equation}
where $R_V$, Eq.~(\ref{RV}), which is a ratio of matrix elements
pertaining to end effects, simplifies at certain special
temperatures, see Eq.~(\ref{specialv}). Note that finite-size
effects are still important for sizes of several thousands of base
pairs (see sections~\ref{DNA} and \ref{finite:size} and
Figs~\ref{fig7} and \ref{fig8}) and are not a purely academic
debate.

Three important correlation lengths can be calculated in the
framework of our model. On the one hand, the Ising correlation
length $\xi_I$, gives access to the typical size of bubbles in the
low temperature regime ($T<T_m$), as well as to the typical size of
unbound regions for $T>T_m$. This quantity is calculated in
Eq.~(\ref{ising}) and assumes a  simplified form at $T_m^\infty$:
$\xi_I (T_m^\infty) \simeq
 \exp[2 J_0(T_m^\infty)]/2 \gg 1$, when $J_0(T_m^\infty) \gg 1$.
On the other hand, one effective chain persistence length,
$\xi^p_{\rm eff,CF} \simeq [\varphi_{U, \infty} / \xi _{U}^p +
\varphi_{B, \infty} / \xi _{B}^p]^{-1}$, provides information on the
short distant behavior of the chain tangent-tangent correlation
function, Eq.~(\ref{tangentcorrNiinflin}), and the other,
$\xi^p_{\rm eff}$, provides information on the typical chain
conformations, in particular its mean-square-radius:
\begin{equation}
\langle {\bf R}^2 \rangle \simeq 2 a^2 N \xi^p_{\rm eff} \simeq 2
a^2 N (\varphi_{U, \infty} \xi^p_U+\varphi_{B, \infty} \xi^p_B),
\end{equation}
where $a$ is the monomer length (0.34~nm) and the approximation is
valid for very long chains (large $N$). Knowing this quantity is of
primary importance when interpreting data from Tethered Particle or
Tweezer experiments~\cite{pouget,Smith92,pouget2}, Atomic Force
Microscopy~\cite{Wiggins06} or viscosity measurements~\cite{inman}.
The results that we have obtained here for the chain tangent-tangent
correlation function and mean-square-radius are very different from
what one obtains by solving a quenched random rigidity model, where
the local joint rigidity can take on one of two values, $\kappa_1$
and $\kappa_2$, with probability $\varphi_1$ and $\varphi_2 = 1 -
\varphi_1$. In this case there is only one effective correlation
length: $\overline{\langle {\bf t}_i \cdot {\bf t}_{i + r} \rangle}
= e^{-r/\xi^p_{\rm ran}}$ and $ \overline{\langle {\bf R}^2 \rangle
} \simeq 2 a^2 N \xi^p_{\rm ran}$, where $\xi^p_{\rm ran} \equiv
-1/\ln[\varphi_1 e^{-1/\xi_p(\kappa_1)} + \varphi_2
e^{-1/\xi_p(\kappa_2)}]$ and $\xi_p(\kappa)$ is given by
Eq.~(\ref{xi}).

The foregoing analysis makes allowance  for neither solvent entropic
(hydrophobic), nor  electrostatic effects, which might not only
change the actual value of the bare Ising parameters $\tilde J$,
$\tilde K$ and $\tilde \mu$, but also lead to additional entropic
contributions. To go further concerning solvent entropic effects
would require molecular dynamics simulations, which  is beyond the
scope of the present approach. The electrostatic effects in DNA
melting are two-fold: an entropic contribution arising from the
difference in electrostatic energy between states $U$ and $B$ and an
an enthalpic contribution arising from  counterion release. At
equilibrium the two effects partially compensate and the remaining
contribution is, using a simple thermodynamic approach in the low
salt limit~\cite{manning2,korolev}, approximately equal to $\Delta
G_{\rm el}=k_BT \ell_b (\tau_B-\tau_U)\ln(I/I_0)$, where $I$ is the
ionic strength ($I_0=1$~ M),  $\ell_b=e^2/(4\pi\epsilon k_BT)$ the
Bjerrum length ($e$ is the elementary charge and $\epsilon$ the
water dielectric permittivity) and $\tau_U$ and $\tau_B$ are the
linear charge densities of the pure U and B chains, respectively.
Although this is an approximate result and the determination of
$\tau_B-\tau_U$ remains controversial, these two  contributions
should be included in a more refined model. A natural extension
emerging from this study concerns the ionic strength dependence of
DNA melting profiles and effective persistence lengths.

The current rapid development of force experiments in magnetic or
optical tweezer traps~\cite{Smith92}, ``Tethered Particle Motion''
experiments~\cite{pouget} or even more recently atomic force
microscopy ones~\cite{Wiggins06}, presents a formidable opportunity
to investigate directly the elastic properties of DNA strands as a
function of temperature, salt concentration and  length $N$. Indeed,
despite the pioneering work by Blake and Delcourt~\cite{blake}, a
systematic experimental study of the effect of $N$ while controlling
the nature of chain ends, is lacking, especially for homopolymers.
This would  be a way to discriminate between the different models
and also  shed light on the role of loop entropy, which is neglected
in the present model.

The approach developed here can be extended to bubble dynamics. Very
recent work~\cite{metzler2,metzler1,mukamel} studied the growth of
already nucleated bubbles using the Fokker-Planck equation applied
to the Poland-Scheraga model (i.e., an  effective Ising model
including loop entropy). The agreement with  experimental results
obtained  by fluorescence correlation spectroscopy is remarkably
good~\cite{metzler2}. The issue of bubble nucleation is, however,
not solved and will be continued to be explored in the near future.
In addition the mutual influence of thermally excited bubbles and
chain flexibility should play an important role  in determining
global DNA conformations and strongly influence  looping
dynamics~\cite{pouget2, finzi}.

\begin{table}[ht]
\begin{center}
\begin{tabular}{|c|l|c|c|}
\hline
Symbol & Quantity & Mathematical definition & Reference \\
\hline $T_m$            & melting temperature &
        $\varphi_{U}(T_m) = \varphi_{B}(T_m) =1/2$ & Secs.~\ref{intro},~\ref{DSM} \\
$ N $            & chain length                 & & Sec.~\ref{DSM} \\
${\rm U}/{\rm B}$            & unbroken/broken base pair        & & Sec.~\ref{DSM} \\
$\varphi_{U,B}$  & fractions of U's and B's     &
                                                & Eq.~(\ref{phidef}) \\
$ \sigma_i$      & internal degree of freedom (Ising variable)
             & $\pm 1$ & Sec.~\ref{DSM} \\
${\bf{t}}_i$     & chain unit tangent vector (Heisenberg variable)
             & $\|{\bf{t}}_i\| = 1$ & Sec.~\ref{DSM} \\
$\tilde{\kappa}_{i,i+1}$ & local chain bending rigidity (coupling)
             & $\kappa_U, \kappa_B$ or $\kappa_{UB}$ & Sec.~\ref{DSM} \\
$\tilde{J}$  & half-energy of a domain wall & & Sec.~\ref{DSM} \\
$\tilde{K}$  & difference in stacking energy between ds and ssDNA
                                                & & Sec.~\ref{DSM} \\
$\tilde{\mu}$ & half-energy required to open a base-pair
                                                & & Sec.~\ref{DSM} \\
$\kappa,J,K,\mu,\ldots$
                 & adimensional energies (in units of $k_BT$)
                 & $\kappa = \beta \tilde{\kappa}$, \ldots & Sec.~\ref{DSM} \\
$ G_0(\kappa)$   & free energy of a single joint of rigidity
$\kappa$
           & $\kappa - \ln[\sinh(\kappa)/\kappa]$ & Sec.~\ref{DSM} \\
$ J_0$       & renormalized (effective) Ising parameter $J$
           & $J - \frac14 [G_0(\kappa_U) + G_0(\kappa_B) - 2G_0(\kappa_{UB})]$
           & Eq.~(\ref{J0}) \\
$ K_0$       & renormalized (effective) Ising parameters $K$
           & $K-\frac12 [G_0(\kappa_U) - G_0(\kappa_B)] $& Eq.~(\ref{K0}) \\
$L_0$            & effective chemical potential to open an
                   interior base pair \ & $\mu + K_0$ & Sec.~\ref{DSM} \\
$\langle c \rangle_\infty$ & infinite size Ising ``magnetization''
            & $\sinh(L_0)/[ \sinh^2(L_0) + e^{-4J_0}]^{1/2}$
            & Eq.~(\ref{c})\\

$\varphi_{U, \infty}$, $\varphi_{B, \infty}$ &  fraction of unbroken
(U) and broken (B) bonds ($N\to\infty$)
            & $[1 \pm \langle c \rangle_\infty]/2$
            & Sec.~\ref{solIsing}   \\

$\varphi_{B, i}$  &  site $i$ bonding opening probability (melting
map)
            & $[1 - \langle \sigma_i \rangle]/2$
            & Eq.~(\ref{phib})   \\

$T_m^{\infty}$ &infinite size ($N \rightarrow \infty$) melting
temperature
               & $L_0(T_m^{\infty})=0$ & Sec.~\ref{DSM} \\
$\Delta T_m^\infty$ & transition width ($N \rightarrow \infty$)
             & \; $2|\partial\langle c \rangle_\infty
               /\partial T|^{-1}$ at ${T_m^{\infty}}$ \& Figs.~\ref{fig5}
and \ref{fig6} \;
                    & Eq.~(\ref{deltaTm}) \\
$\langle \sigma_{i+r} \sigma_i \rangle$ & Ising correlation function
                 &        & Sec.~\ref{corr} \\
$\xi_{I}$      & Ising correlation length
               & $e^{2J_0}/2$ at ${T_m^{\infty}}$ \& Fig.~\ref{fig4}
               & Sec.~\ref{corr} \\
$\langle {\bf{t}}_{i+r} \cdot {\bf{t}}_i \rangle$ & chain
correlation function
               & Fig.~\ref{fig3}(b) & Sec.~\ref{TM}\\
$\xi^p_{1,+},\xi^p_{1,-}$ & persistence lengths & Fig.~\ref{fig4}
               & Eq.~(\ref{tangentcorrNiinf}) \\
$\xi^p_{{\rm eff}}$ & effective persistence length & Fig.~\ref{fig4}
               & Sec.~\ref{Rg} \\
$R$            & chain mean square (or gyration) radius
               & $\langle {\bf R}^2\rangle^{1/2}$ & Secs.~\ref{DSM},\ref{Rg}\\
$T^*$            & crossover temperature for finite chains \ &
$\partial \varphi_B/\partial N |_{T^*}=0$ & Sec.~\ref{finite:size} \\
\hline
\end{tabular}
\end{center}
\caption{Index of the main symbols used throughout the paper, with
their mathematical definition, and reference.}\label{table1}
\end{table}

\end{document}